\documentclass[a4paper,11pt]{article}
\pdfoutput=1
\usepackage{vmargin}
\setmarginsrb           { 2.5cm}  
                        { 1.5cm}  
                        { 2.5cm}  
                        { 1.5cm}  
                        {  20pt}  
                        {0.25in}  
                        {   9pt}  
                        { 0.3in}  
\usepackage[square, numbers, comma, sort&compress]{natbib}  
\usepackage{color}
\usepackage{amsmath,amsfonts,amssymb,amscd,amsthm,xspace}
\usepackage{graphicx}
\usepackage[centerlast,small,sc]{caption}
\usepackage[pdfpagemode={UseOutlines},bookmarks=true,bookmarksopen=true,bookmarksopenlevel=0,bookmarksnumbered=true,hypertexnames=false,colorlinks,linkcolor={blue},citecolor={blue},urlcolor={blue},pdfstartview={FitV},unicode,breaklinks=true]{hyperref}
\renewcommand{\thesection}{\arabic{part}.\arabic{section}}
\renewcommand{\thesubsection}{\arabic{part}.\arabic{section}.\arabic{subsection}}
\makeatletter
    \renewcommand{\theequation}{%
    \arabic{part}.\arabic{section}.\arabic{equation}}
    \@addtoreset{equation}{section}
\makeatother
\begin{document}
\title  {Regularization and conformal transformations\\of the power spectrum\\in general single field inflation}
\author{Yukari Nakanishi\footnote{nakanishi@het.phys.sci.osaka-u.ac.jp}\\{\small Department of Physics, Osaka University, Toyonaka, Osaka 560-0043, Japan}}
\date       {}

\maketitle




\begin{center}{\small Received February 3, 2015}\end{center}
\vfil
    \abstract{
    The regularization of the CMB power spectrum is an important issue of cosmology.
    Most approaches assume that there is no need to regularize the power spectrum, while Parker advocated the new regularization approach for the power spectrum in 2007 \cite{Parker:2007-1}: the adiabatic regularization, which was originally developed for particle creation in curved spacetime (see the review \cite{Parker:2012-1}). This thesis focuses on this issue, especially concerning adiabatic subtraction terms.
    The subtraction terms for minimally coupled slow-roll inflation are well known (see e.g. \cite{Parker:2007-1}).
    We extend the view to more generic inflation models, and derive the subtraction terms for k-inflation models.
    Via the method of Urakawa--Starobinsky \cite{Urakawa:2009-1}, we consider the time development of the subtraction term at late times. We also consider the non-minimally coupled case, and show that the adiabatic regularization is independent of the frame: Jordan or Einstein frames.
    
    }

\clearpage  


\begin{center}{\Large Acknowledgements}\end{center}
I would like to express my deep appreciation to Prof. Yutaka Hosotani, who provided tremendous support to my studies.
I also greatly thank Prof. Takahiro Kubota and Dr. Wade Naylor, whose meticulous feedback and valuable comments greatly helped me.
Dr. Naylor gave me such an interesting theme and assisted to improve my English.
I also benefited from discussions with my collaborator Mr. Allan L. Alinea and would like to show my gratitude to him as well.
\vspace{0.5cm}

\noindent This report was originally written for the author's Master's Thesis at the Particle Physics Theory Group, Department of Physics, Graduate School of Science, Osaka University. This version has added some extra material (Appendix \ref{chap:sub-dev-after} and \ref{chap:jordan-frame}) for the arXiv preprint server version.
\clearpage  


\tableofcontents  





\newpage\part{Introduction}
The inflationary paradigm \cite{Guth:1981-1,Sato:1981-1,Linde:1982-1,Albrecht:1982-1} is a fascinating approach to solving various cosmological problems.
Although inflation can describe the solutions naturally, we do not know how it occurs, or even whether or not it really occurred.

To examine inflation, we have a tool to see the early universe: the cosmic microwave background (CMB) radiation.
The CMB consists of photons which became free at early times (the last scattering surface) and now fill the observable universe.
We can observe the temperature and the polarization, and we know that the temperature of these photons from each direction is almost identical.

However, tiny fluctuations in the temperature and the polarization have a large amount of information of the early universe because it is considered that they reflect the primordial perturbations of inflaton or gravitation.
If we can analyze the observation data properly, it means that we can see the universe at the last scattering surface.
Indeed, the CMB is the oldest light that we can observe. To know the details of the universe before the last scattering surface directly, we are forced to detect other signals such as gravitational wave signatures. 

We can infer from the CMB that the early universe went through an inflationary era \cite{Planck:2013-1}.
However, the precision is still not enough to determine which inflation model was realized, and many inflation models have been advocated from 
various theoretical motivations. To compare theoretical inflation models with the observed CMB, the power spectrum which will be shortly explained in section (\ref{sec:powerspectrum}) is used. Therefore the method used to estimate the theoretical power spectrum is an important matter of cosmology.

Although the amplitude of the perturbation which is the integrand of the power spectrum has divergences (before regularization), usually it has 
been assumed that regularization is not necessary\footnote{%
Recently, a review of the problem \cite{BasteroGil:2013-1} has even claimed that there is no need to regularize the power spectrum from the standpoint of observations (although interacting theories would still require a regularization according to \cite{BasteroGil:2013-1}, see also \cite{delRio:2014-1}).
}.
However, since the paper of Parker \cite{Parker:2007-1} quite a lot of discussion has arisen concerning this issue.
He applied the adiabatic regularization method, which is a regularization method used in quantum field theory in curved spacetime, to the power spectrum.

The purpose of this thesis is to review the adiabatic regularization for the power spectrum and to expand it to more general cases.
In the rest ot this part, Part 1, we review cosmological perturbation theory and we also review adiabatic regularization and k-inflation in the first half of Part 2. 
Then we expand our interest to non-minimally coupled k-inflation, and derive the adiabatic subtraction terms for the model in the last part of Part 2.
In Part 3, we discuss the issue of the physicality Einstein and Jordan frames that are related by a conformal transformation, where of relevance to this thesis we show the conformal equivalence of the adiabatic subtraction terms in each frame. In Part 4 we conclude and mention future work.
We include detailed calculations in Appendices: In Appendix A, we calculate Hubble flow functions in detail. In Appendices B and C, the time development of the adiabatic subtraction terms is discussed in some specific models. Finally, in Appendix D, we try to express the subtraction terms in the Jordan frame and discuss the generality of the derived subtraction terms.

\section{ Cosmological perturbation theory} 
\subsection{The Friedmann equations}
Observationally, the universe is nearly homogeneous and isotropic and 
hence perturbation theory is usually considered around such a background.
In this thesis, we go up to linear order in perturbations, although in principle higher order could be considered.

\subsubsection*{The background metric}
The homogeneous and isotropic background universe is written by the Friedmann--Lema\^{i}tre--Robertson--Walker (FLRW) metric.
\begin{eqnarray}
ds^2&=&dt^2-a^2(t)d\vec{x}^2\nonumber\\
&=& dt^2 -a^2 (t) \left[ \frac{dr^2}{1-Kr^2} +r^2 (d\theta^2 +\sin^2 \theta d\phi^2) \right]
\label{eq:FLRW-t}
\end{eqnarray}
where $K$ is spatial curvature. $a(t)$ is a scale factor which denotes the size of the universe at specific time. In this thesis, only the spatially flat ($K=0$) case will be considered.

To simplify the calculations, conformal time $\eta$ defined as
\begin{equation}
\eta =\int^t \frac{dt'}{a(t')}
\end{equation}
is also used. The metric is written in terms of $\eta$ as follows
\begin{equation}
ds^2=a^2(\eta)\left[d\eta^2-d\vec{x}^2\right]~.
\label{eq:FLRW}
\end{equation}
Both proper time $t$ and conformal time $\eta$ are used in this thesis.
The derivatives is written by
\begin{equation}
\dot{f}(t)\equiv\frac{d}{dt}f(t),\hspace{1em}f'(\eta)\equiv\frac{d}{d\eta}f(\eta)~.
\end{equation}
However, we use Eq. (\ref{eq:FLRW}) as the background metric unless otherwise specified.

\subsubsection*{Metric perturbations}
Then let us consider the perturbations.
The metric perturbations has three modes.
1) Scalar perturbation $\phi(\eta,\vec{x}),\psi(\eta,\vec{x}),B(\eta,\vec{x}),\tilde{E}(\eta,\vec{x})$.
2) Vector perurbation $F(\eta,\vec{x})_i ,S(\eta,\vec{x})_i$.
3) Tensor perturbation $h(\eta,\vec{x})_{ij}$.
\begin{eqnarray}
\delta g_{00}(x)&=&2a^2\phi\\
\delta g_{0i}(x)&=&a^2(B,_i +S_i)\\
\delta g_{ij}(x)&=&a^2(2\psi\delta_{ij}+2\tilde{E},_{ij}+F_{i,j}+F_{j,i}+h_{ij})
\label{eq:metricP3}
\end{eqnarray}

If we pick all scalar modes only, the line element is written as
\begin{equation}
ds^2=a^2(\eta)\left[\left(1+2\phi\right)d\eta^2+2B,_i d\eta dx^i -\left\{\left(1-2\psi\right)\delta_{ij} -2\tilde{E},_{ij}\right\}dx^i dx^j \right]~.
\label{eq:perturbedm}
\end{equation}
The above metric will eventually be used to derive an equation of motion for the inflation perturbations, known as the Muhkanov--Sasaki equation.

\subsubsection*{Hydrodynamical perturbations}
Classically, the universe is homogeneous and isotropic, and matter can be approximated by a  perfect fluid.

The energy-momentum tensor of perfect fluid is given by following form.
\begin{equation}
T_{\mu\nu}=(E+P)u_{\mu}u_{\nu}-Pg_{\mu\nu}
\label{eq:EMtensor}
\end{equation}
where $E$ is the energy density of the matter and $P$ is the pressure.
$u^{\mu}(x)$ is a four vector which satisfy the classical part $u_{c\mu} =(a,0,0,0)$ and $u_{\mu}u^{\mu}=1$.

In perturbation theory, $E$, $P$ and $u^{\mu}$ also have the spatial-dependent perturbation part.
Let us write the classical part like $E_c$. e.g.) $E(x)=E_c(\eta)+\delta E(\eta,\vec{x})$.
The perturbation of $u_i$ can be decomposed scalar part and vector part, so write them explicitly.
\begin{equation}
\delta u_i =\delta u_{\parallel},_i +\delta u_{\perp i}
\end{equation}
where $\partial^i \delta u_{\perp i} =0$ and $\delta u_{\parallel}$ is the scalar part.

%
The classical part of the energy-momentum tensor becomes
\begin{equation}
T^0_{c\ 0} =E_c ,\hspace{1em}
T^0_{c\ i} =0 ,\hspace{1em}
T^i_{c\ j} = -P_c \delta^i_{\ j}
\end{equation}
and the linear perturbation part (only scalar modes) are given by
\begin{eqnarray}
\delta T^0_{\ 0} &=& \delta E~,\nonumber\\
\delta T^0_{\ i} &=& \frac{1}{a}(E_c +P_c )\delta u_{\parallel},_i ~, \label{eq:EMtensor3}\\
\delta T^i_{\ j} &=& -\delta P\delta^i_{\ j}~.\nonumber
\end{eqnarray}
These equations will be used later on to derive the Muhkanov--Sasaki equation after we have first defined the cosmological background.

\subsubsection*{The Friedmann equations}
Using these metric and energy-momentum tensor, we can get the Friedmann equation from the Einstein equation $R_{\mu\nu}-\frac{1}{2}Rg_{\mu\nu}=T_{\mu\nu}$\footnote{In this thesis $\sqrt{8\pi G}\equiv M_{\text{Pl}}^{-1}$ is normalized so that $M_{\text{Pl}}=1$.}.

The classical part is
\begin{equation}
H^2=\frac{E_c}{3},\hspace{1em}\frac{\ddot{a}}{a}=-\frac{1}{6}(E_c +3P_c)
\label{eq:Friedmann1}
\end{equation}
where $H\equiv\frac{\dot{a}}{a}$.
The continuity equation is also derived
\begin{equation}
\dot{E}_c +3H(E_c +P_c)=0~.
\label{eq:continuity1}
\end{equation}

Or in terms of conformal time,
\begin{equation}
\mathcal{H}^2=\frac{a^2 E_c}{3},\hspace{1em}\frac{a''}{a}-\frac{a'^2}{a^2}=-\frac{a^2}{6}(E_c +3P_c)
\label{eq:Friedmann2}
\end{equation}
\begin{equation}
E'_c +3\mathcal{H}(E_c +P_c)=0
\label{eq:continuity2}
\end{equation}
where $\mathcal{H}\equiv\frac{a'}{a}=aH$.

The equation of scalar perturbation is more complicated because the metric perturbation contains non-physical degrees of freedom. To deal with this problem, we fix the gauge or compose gauge invariant quantities as discussed in the following section.

\subsection{Gauge invariance} 
Gauge fixing can remove the non-physical degrees of freedom in the perturbed metric Eq. (\ref{eq:perturbedm}).
However, the variables in the gauge are not always physical quantity, i.e., they may be not gauge invariant.

In this thesis the conformal-Newtonian gauge will be considered.

\subsubsection*{The conformal-Newtonian gauge}

The conformal-Newtonian (longitudinal) gauge is defined by
\begin{equation}
ds^2=a^2(\eta)\left[\left(1+2\phi\right)d\eta^2-\left(1-2\psi\right)d\vec{x}^2\right]~.
\label{eq:CNgauge}
\end{equation}
i.e., $B=E=0$.

In this gauge, the perturbations of the metric and inflaton are gauge invariant.
Thus we can rewrite metric Eq. (\ref{eq:CNgauge}) or equations in conformal-Newtonian gauge in terms of gauge invariant variables. Using the gauge invariant perturbations $\Phi$ and $\Psi$, the metric becomes
\begin{equation}
ds^2=a^2(\eta)\left[\left(1+2\Phi\right)d\eta^2-\left(1-2\Psi\right)d\vec{x}^2\right]~.
\label{eq:CNgauge-i}
\end{equation}
$\Phi$ corresponds to the Newtonian potential.
There seems to be two degrees of freedom for the scalar mode.
Actually, $\Phi$ and $\Psi$ are related via the Einstein equation and not independent.
In minimal coupling case, these two are identical $\Phi=\Psi$.

For other perturbations, the overlines will be used to express gauge invariant variables. e.g., $\delta\varphi(\eta,\vec{x})\rightarrow\overline{\delta\varphi}(\eta,\vec{x})$.

\subsubsection*{The comoving gauge}
The comoving gauge which is another gauge fixing way is also considered in some references.
We will not use it in this thesis, but briefly explain it.

The comoving gauge is defined by the condition
\begin{equation}
\phi=0,\ \partial^i \delta u_{\parallel} =0~.
\end{equation}
The scalar part of the metric becomes
\begin{equation}
ds^2 = a^2 (\eta)\left[d\eta^2 -e^{2\mathcal{R}}\delta_{ij}dx^i dx^j \right]
\end{equation}
where $\mathcal{R}$ is the comoving curvature perturbation which is a gauge-invariant quantity composed by scalar perturbations in the metric and the inflaton.

This gauge is particularly well suited to performing calculations using the ADM formalism \cite{Arnowitt:1959-1} in cosmology, e.g., see \cite{Maldacena:2002-1,Baumann:2009-1}.

\subsection{Mukhanov--Sasaki equation}

After gauge fixing, the first order Einstein equations can be solved.
We can calculate it in conformal-Newtonian gauge, and then replace the perturbations by the gauge invariant ones.

In the minimal coupling case, we can combine the Einstein equations and get the Mukhanov--Sasaki equation describing the behavior of the comoving curvature perturbation $\mathcal{R}$.
\begin{equation}
v''_k +\left(c_s^2 k^2 - \frac{z''}{z}\right)v_k=0
\label{MSeq}
\end{equation}
where
\begin{equation}
v_k\equiv z\mathcal{R}_k,\hspace{1em}
z\equiv\frac{a^2 \sqrt{E+P}}{c_s \mathcal{H}},\hspace{1em}
c_s^2\equiv\frac{\partial P}{\partial E}~.
\label{eq:zdef}
\end{equation}

The comoving curvature perturbation $\mathcal{R}\equiv\Psi-\frac{\mathcal{H}}{a^2}\delta u_{\parallel}$ \cite{Baumann:2009-1} is a gauge invariant scalar which is conserved and is equal to the curvature perturbation on uniform-density hypersurfaces $\zeta$ on superhorizon scales.

Finally, we should mention that $c_s$ is the "sound speed," the propagation speed of the perturbation.
In models which have canonical kinetic terms we have $c_s=1$.

\section{ The power spectrum\label{sec:powerspectrum}}
\subsection{Quantization}

In this section, the definition and the physical property of the power spectrum of the CMB temperature is introduced. 

First, we consider how to quantize the Mukhanov--Sasaki variable $v(x)$ to $\hat{v}(x)$.
At first, define the canonical momentum $\hat{\pi}$,
\begin{equation}
\hat{\pi}\equiv\frac{\partial\mathcal{L}}{\partial\hat{v}}
\end{equation}
where $\mathcal{L}$ is the Lagrangian. We then assume that the quantized variable $\hat{v}$ also obeys Mukhanov--Sasaki equation.

We now require the canonical commutation relation on $\hat{v}$ and $\hat{\pi}$:
\begin{equation}
[\hat{v}(\eta,\vec{x}),\hat{v}(\eta,\vec{y})]=[\hat{\pi}(\eta,\vec{x}),\hat{\pi}(\eta,\vec{y})] =0
\end{equation}
\begin{equation}
[\hat{v}(\eta,\vec{x}),\hat{\pi}(\eta,\vec{y})] = i\delta (\vec{x}-\vec{y})
\end{equation}

Let us consider the Fourier components
\begin{equation}
\hat{v}(\eta,\vec{x})=\int\frac{d^3 k}{(2\pi)^{\frac{3}{2}}}\left[
A_{\vec{k}}v_k (\eta)e^{i\vec{k}\cdot\vec{x}}+
A_{\vec{k}}^{\dagger} v_k^* (\eta)e^{-i\vec{k}\cdot\vec{x}}
\right]
\label{eq:Fourierv}
\end{equation}
with the normalization
\begin{equation}
v_k^* v'_k -v_k^{*} {}' v_k =-i~.
\end{equation}
Then $A_{\vec{k}}$ and $A_{\vec{k}}^{\dagger}$ have the canonical commutation relation
\begin{equation}
[A_{\vec{k}},A_{\vec{k}'}]=[A_{\vec{k}}^{\dagger},A_{\vec{k}'}^{\dagger}]=0,\hspace{1em}
[A_{\vec{k}},A_{\vec{k}'}^{\dagger}]=\delta (\vec{k}-\vec{k}')~.
\label{eq:Acom}
\end{equation}

The power spectrum of two point correlation function for scalar perturbation $\mathcal{P}_{\mathcal{R}}(k)$ is defined by
\begin{equation}
\langle\mathcal{R}_{\vec{k}}\mathcal{R}_{\vec{k}'}\rangle\equiv\delta(\vec{k}+\vec{k}')\mathcal{P}_{\mathcal{R}}(k)~.
\label{eq:Pdef}
\end{equation}

Using $v_k\equiv z\mathcal{R}_k$ and Eqs. (\ref{eq:Fourierv})--(\ref{eq:Acom}):
\begin{eqnarray}
\langle\mathcal{R}(\eta,\vec{x})\mathcal{R}(\eta,\vec{y})\rangle
&=&\frac{1}{z^2}\langle\hat{v}(\eta,\vec{x})\hat{v}(\eta,\vec{y})\rangle\nonumber\\
&=&\frac{1}{z^2}\frac{1}{(2\pi)^3}\int d^3 k |v_k|^2 e^{i\vec{k}\cdot(\vec{x}-\vec{y})}\nonumber\\
&=&\int\frac{dk}{k}\frac{\sin{(k|\vec{x}-\vec{y}|)}}{k|\vec{x}-\vec{y}|}\frac{k^3}{2\pi^2} |\mathcal{R}_k|^2
\end{eqnarray}
If we set $\vec{x}=\vec{y}$,
\begin{equation}
\langle\mathcal{R}^2 (\eta,\vec{x})\rangle=\int\frac{dk}{k}\frac{k^3}{2\pi^2} |\mathcal{R}_k|^2~.
\label{eq:powerspectrum}
\end{equation}

On the other hand, from Eq. (\ref{eq:Pdef}):
\begin{equation}
\langle\mathcal{R}^2 (\eta,\vec{x})\rangle=\frac{1}{2\pi^2}\int dk\ k^2 \mathcal{P}_{\mathcal{R}}(k)~.
\end{equation}
\begin{equation}
\therefore \mathcal{P}_{\mathcal{R}}(k) = |\mathcal{R}_k|^2
\end{equation}

The integrand of Eq. (\ref{eq:powerspectrum}) $\Delta^2_{\mathcal{R}}(k)\equiv\frac{k^3}{2\pi^2} \mathcal{P}_{\mathcal{R}}(k)$ is called the dimensionless power spectrum. In this thesis, we will sometimes call $\Delta^2$ the power spectrum.
However, we make it clear which we indicate from the context when the distinction is needed.

\subsection{Behavior of the power spectrum}

Given that the comoving curvature perturbation $\mathcal{R}$ is conserved after the Hubble horizon crossing, the power spectrum also time-independent after crossing. This is the reason why we are especially interested in large scales.
The modes in large scales exit early in the history of the universe, and they keep more information of primordial perturbations than the modes in small scales.

Is it known that the dimensionless power spectrum of the scalar perturbation in de Sitter spacetime can be exactly solved and it does not depend on $k$. Moreover, the observations support the nearly $k$-independent power spectra\cite{Planck:2013-1}.
If the dimensionless power spectrum can be expressed (or approximated) by power series, the integral Eq. (\ref{eq:powerspectrum}), which is the variance of the perturbation, has the divergences on UV scales.
Thus we need to regularize the power spectrum 
in theoretical arguments.

In next part, we will discuss the adiabatic regularization which is a way to regularize the power spectrum.
We will see the behavior the adiabatic subtraction term in the latter half of Part 2.
\newpage\part{Regularization of the power spectrum}
\setcounter{section}{0}

To obtain the finite expression of power spectrum, adiabatic regularization\cite{Parker:2007-1,ParkerToms200908}, which is one of regularization schemes of quantum field theory in curved spacetime, is considered.
In this regularization, we perform WKB-like expansion of the solution of perturbations, and subtract the isolated divergence part.
Then we regard the regularized power spectrum as  the corresponding one to the observable power spectrum.

In this chapter, we review the adiabatic regularization method.
And then we apply it to the power spectrum for k-inflation model.

\section{ Adiabatic regularization\label{sec:adiabatic}} 
Let us consider a scalar field $\phi(x)$ obeying the following equation of motion\footnote{We formulate the adiabatic regularization method using the conformal time for some convenience. The original formulation \cite{Parker:2007-1,ParkerToms200908} uses the proper time, and the both subtraction terms are equivalent. We can obtain the subtraction term in terms of proper time by using either the differential equation by proper time or by the variable transformation of the subtraction term derived by Eq. (\ref{eq:genEOM}).}.
\begin{equation}
\phi''(x)+f(\eta)\phi'(x)+\left[g(\eta)\partial_i \partial^i +h(\eta) \right]\phi(x)=0
\label{eq:genEOM}
\end{equation}
where $f(\eta)$, $g(\eta)$ and $h(\eta)$ are functions of time and these can be derived from the given metric and Lagrangian\footnote{We assume that the potential is not steep.}.

In this case, the coefficient functions from the metric depends on only time because we are considering up to first order of perturbations.
For the power spectra, we will consider the equation of motion of the perturbation part only.
The background metric is 
Eq. (\ref{eq:CNgauge-i}), which has 
the both dependence of time and spatial coordinates. However, the dependence on the spatial coordinates appears as the perturbations.
Therefore the spatial-dependence of the metric appears the second or higher order perturbation parts, and we can use the Eq. (\ref{eq:genEOM}) in our discussion without loss of generality.

Consider the Fourier transformation:
\begin{equation}
\phi(x)=\int \frac{d^3 k}{(2\pi)^{\frac{3}{2}}} \left(A_{\vec{k}}\phi_{\vec{k}}(x)+A_{\vec{k}}^{\dagger}\phi_{\vec{k}}^{*}(x)\right)
\label{eq:fourier}
\end{equation}
The following normalization of $\phi$ is also imposed.
\begin{equation}
\phi_k^* \phi'_k -\phi_k^{*} {}' \phi_k =-i
\end{equation}

If we define $\chi_k (\eta)$ as
\begin{equation}
\phi_{\vec{k}}(x)\equiv\chi_k (\eta)\exp{\left(i\vec{k}\cdot\vec{x}-\frac{1}{2}\int f(\eta)d\eta\right)}~,
\end{equation}
$\chi_k (\eta)$ obeys the following equation.
\begin{equation}
\chi''_k (\eta)+\Omega^2(\eta)\chi_k (\eta)=0
\label{eq:chiEOM}
\end{equation}
where
\begin{equation}
\Omega^2 (\eta)\equiv-g(\eta)k^2 +h(\eta)-\frac{1}{4}f^2 (\eta)-\frac{1}{2}f'(\eta)
\label{eq:Omega}
\end{equation}
This is an equation of motion of a harmonic oscillator if we regard $\Omega(\eta)$ as the frequency.
Indeed, in the case of a massive scalar field in the Minkowski spacetime,
\begin{equation}
f(\eta)=0,~g(\eta)=-1,~h(\eta)=m_{\phi}^2
\end{equation}
and Eq. (\ref{eq:chiEOM}) becomes
\begin{equation}
\chi''_k (\eta)+\left(k^2 +m_{\phi}^2 \right) \chi_k (\eta)=0~.
\end{equation}
$k^2 +m_{\phi}^2 \equiv\omega_k^2$ is the square of the frequency and the solution is
\begin{equation}
\chi_k (\eta) \propto \exp{\left(-i\omega_k \eta \right)}~.
\end{equation}

In a curved spacetime
, the square of effective frequency $\omega^2_k (\eta) \equiv-g(\eta)k^2 +h(\eta)$ depends on time and the coefficient function $f(\eta)$ from the Christoffel symbol in covariant derivative exists. Therefore we perform a WKB-like (adiabatic) expansion for the solution.

Introducing a fictitious parameter $T$ in the metric.
\begin{equation}
g_{\mu\nu} (x) \rightarrow g_{\mu\nu} (x/T)
\end{equation}
The order of differentiation of the metric becomes the adiabatic order of the term. We now require the adiabatic condition.
i.e., In lowest adiabatic order, $\chi_k$ should have the form
\begin{equation}
\chi_k \sim \frac{1}{\sqrt{2 \omega_k (\eta)}} \exp{\left(-i\int^{\eta} \omega_k (\eta')d\eta'\right)}~.
\end{equation}
Under the adiabatic condition, we can get uniquely the adiabatic expansion of the solution $\chi_{\vec{k}}$, and the solution is schematically written by
\begin{equation}
\chi_k (\eta) = \frac{1}{\sqrt{2W_{2n}(\eta)}}\exp{\left(-i\int^{\eta}W_{2n}(\eta')d\eta'\right)}+\mathcal{O}(T^{-2n-2})
\label{eq:yuugen}
\end{equation}
up to $2n$th adiabatic order. $W_{2n}$ is obtained later by the effective frequency of the equation of motion each order.

$\phi_{\vec{k}}$ become the complete orthogonal set of solutions, and $A_{\vec{k}}$ and $A^{\dagger}_{\vec{k}}$ becomes annihilation and creation operators after requiring the canonical commutation relations
\begin{equation}
[A_{\vec{k}},A_{\vec{k}'}]=[A_{\vec{k}}^{\dagger},A_{\vec{k}'}^{\dagger}]=0,\hspace{1em}
[A_{\vec{k}},A_{\vec{k}'}^{\dagger}]=\delta (\vec{k}-\vec{k}'),
\end{equation}
and an adiabatic vacuum state $|0\rangle$ is constructed.
\begin{equation}
A_{\vec{k}}|0\rangle=0\hspace{3em}(\text{for all }\vec{k})
\end{equation}

Neglect the higher order terms in Eq. (\ref{eq:yuugen}).
i.e., Consider the $2n$th adiabatic order solution.
\begin{equation}
\chi_k^{(2n)} (\eta) = \frac{1}{\sqrt{2W_{2n}(\eta)}}\exp{\left(-i\int^{\eta}W_{2n}(\eta')d\eta'\right)}
\end{equation}
$\chi_k^{(2n)} (\eta)$ obeys the following equation.
\begin{equation}
\chi''_k {}^{(2n)} +\left[W_{2n}^2
-W_{2n}^{\frac{1}{2}}\left(W_{2n}^{-\frac{1}{2}}\right)''\right]\chi_k^{(2n)}=0
\label{eq:2nth}
\end{equation}

Rewrite Eq. (\ref{eq:chiEOM}) to
\begin{equation}
\chi''_k +\left[W^2 -W^{\frac{1}{2}}\left(W^{-\frac{1}{2}}\right)''\right]\chi_k
=\left[W^2 -W^{\frac{1}{2}}\left(W^{-\frac{1}{2}}\right)''-\Omega^2\right]\chi_k~.
\end{equation}
In $2n$th adiabatic order, we can choose $W$ so that the left hand side vanishes by using Eq. (\ref{eq:2nth}).
\begin{equation}
0=\left[W^2 -W^{\frac{1}{2}}\left(W^{-\frac{1}{2}}\right)''-\Omega^2\right]\chi_k
\label{eq:higher-rhs}
\end{equation}
i.e., The right hand side is higher adiabatic order than $2n$th order. We can get from Eq. (\ref{eq:higher-rhs})
\begin{equation}
W^2 =\Omega^2 +W^{\frac{1}{2}}\left(W^{-\frac{1}{2}}\right)''
\label{eq:Weq}
\end{equation}
where $W$ in arbitrary order.

Note that $\Omega^2$ of Eq. (\ref{eq:Omega}) can be decomposed to the terms of 0th adiabatic order and 2nd order because $f(\eta)$ is 1st adiabatic order.
Therefore we can write $\Omega^2$ in terms of the 0th adiabatic order part $\omega_k^2$ and the 2nd order part $\sigma\equiv-\frac{1}{4}f^2 (\eta)-\frac{1}{2}f'(\eta)$.
\begin{equation}
\Omega^2 =\omega_k^2 +\sigma
\end{equation}
$W$ is decomposed as following by using the Eq. (\ref{eq:Weq}).
\begin{eqnarray}
W&=&W^{(0)}+W^{(1)}+W^{(2)}+\cdots\\
W^{(0)}&=&\omega_k\nonumber\\
W^{(1)}&=&0\nonumber\\
W^{(2)}&=&\frac{1}{2}\omega_k^{-\frac{1}{2}}\left(\omega_k^{-\frac{1}{2}}\right)'' +\frac{1}{2}\omega_k^{-1}\sigma\nonumber
\end{eqnarray}

The unregularized (``bare'') two point function of $\phi$ becomes
\begin{eqnarray}
\langle0|\phi(x)\phi(\tilde{x})|0\rangle &=&
\frac{1}{(2\pi)^3}\int d^3\! kd^3\! k' \phi_{\vec{k}}(x)\phi_{\vec{k}'}^* (\tilde{x})\delta (\vec{k}-\vec{k}')\nonumber\\
&=&\frac{1}{(2\pi)^3}\int d^3\! k \phi_{\vec{k}}(x)\phi_{\vec{k}}^* (\tilde{x})\nonumber\\
&=&\frac{1}{(2\pi)^3}\frac{1}{\exp{\left(\int\! f(\eta)d\eta\right)}}\int d^3\! k e^{i\vec{k}\cdot (\vec{x}-\vec{\tilde{x}})} \chi_k (x)\chi_k^* (\tilde{x})
\end{eqnarray}
and take limit $\tilde{x}\rightarrow x$.
\begin{eqnarray}
\langle0|\phi^2 (x)|0\rangle&=&\frac{1}{(2\pi)^3}\frac{1}{\exp{\left(\int\! f(\eta)d\eta\right)}}\int d^3\! k |\chi_k (x)|^2\nonumber\\
&=& \frac{1}{2\pi^2}\frac{1}{\exp{\left(\int\! f(\eta)d\eta\right)}}\int dk\ k^2 |\chi_k (x)|^2
\end{eqnarray}

To regularize this, we need $W^{-1}$ up to second order.
\begin{equation}
W=\frac{1}{W^{(0)}+W^{(1)}+W^{(2)}+\cdots}=W^{-1(0)}+W^{-1(1)}+W^{-1(2)}+\cdots
\end{equation}
\begin{equation}
W^{-1(0)}=\omega_k^{-1},\hspace{1em}W^{-1(1)}=0,\hspace{1em}W^{-1(2)}=-\omega_k^{-2}W^{(2)}
\label{eq:W}
\end{equation}

Now the divergent part is uniquely isolated by the adiabatic expansion.
Because only $W^{-1(0)}\propto k^{-1}$ and $W^{-1(2)}\propto k^{-n}\ (n\geq3)$ contain the divergent parts, we subtract these two terms and define the physical two point function (minimal subtraction scheme)
\begin{equation}
\langle0|\phi^2 (x)|0\rangle_{\text{phys}}
\equiv\frac{1}{2\pi^2}\frac{1}{\exp{\left(\int\! f(\eta)d\eta\right)}}\int dk\ k^2 \left[|\chi_k (\eta)|^2-\frac{1}{2}\left\{W^{-1(0)}+W^{-1(2)}\right\}\right]~.
\end{equation}
From Eq. (\ref{eq:W}), it is found that the 0th order subtraction term $W^{-1(0)}$ corresponds to the zero-point energy and its time dependence is due to the expansion of the universe and/or the non-canonical kinetic term.
The second order term does not exist in Minkowski spacetime.

We can use this method to regularize the power spectrum.
The calculation for usual slow-roll inflation model has done, and it is shown that the subtraction terms for the scalar perturbation and the tensor perturbation become sufficiently negligible after the horizon crossing \cite{Urakawa:2009-1} (see also \cite{Agullo:2010-1}).

While the adiabatic subtraction terms for slow-roll inflation model is studied very well, for other inflation models is not.
Hence we generalize it for k-inflation model which has more generic Lagrangian than slow-roll model.

\section{ Minimally coupled k-inflation} 
\subsection{k-inflation}
k-inflation (kinetically driven inflation) is a inflation model advocated by \cite{ArmendarizPicon:1999-1,Garriga:1999-1}.
In this model, the inflaton have non-canonical kinetic terms and they make it possible that the universe expands exponentially.
We will review briefly k-inflation model in this section.

The Lagrangian of inflaton in k-inflation model given by
\begin{equation}
\mathcal{L}_{\varphi}=P(\varphi,X)=-V(\varphi)+K(\varphi)X+L(\varphi)X^2+\cdots
\label{eq:k-lag}
\end{equation}
where $\varphi$ is inflaton and  $X\equiv\frac{1}{2}\partial_{\mu}\varphi\partial^{\mu}\varphi$.
However, the explicit form of the Lagrangian will not be specified except Appendix \ref{chap:sub-dev}, and the general form $P(\varphi,X)$ will be used.

The action in minimally coupled case is
\begin{equation}
S=\frac{1}{2}\int d^4 x\sqrt{-g}\left[-R+P(\varphi,X) \right]~.
\end{equation}

The energy-momentum tensor under the perfect fluid approximation is
\begin{equation}
T_{\mu\nu}=(E+P)u_{\mu}u_{\nu}-Pg_{\mu\nu}~.
\tag{\ref{eq:EMtensor}}
\end{equation}
Using this expression and the definition of the energy-momentum tensor
\begin{equation}
T_{\mu\nu}\equiv\frac{2}{\sqrt{-g}}\frac{\delta S}{\delta g^{\mu\nu}}
=P,_X \partial_{\mu}\varphi \partial_{\nu}\varphi-Pg_{\mu\nu}~,
\end{equation}
the energy $E$ is obtained as $E=2XP,_{X}-P$ where the four-velocity $u_{\mu}\equiv\frac{\partial_{\mu}\varphi}{\sqrt{2X}}$.

To derive the adiabatic subtraction terms in following sections, it is need to obtain the expression of the sound speed.
The sound speed of k-inflation is given by
\begin{equation}
c_s^2 \equiv\frac{P,_X}{E,_X}=\frac{P,_X}{2XP,_{XX}+P,_X}~.
\end{equation}
$c_s^2$ can be negative and $c_s$ can exceed $1$ by definition.
However, $c_s^2 \geq0$ is required in general for stability \cite{ArmendarizPicon:1999-1} and $c_s$ does not exceed the speed of light as long as the Lagrangian does not contain the negative power of $X$.

Also $z$ is expressed in terms of $P$ and $X$,
\begin{equation}
z\equiv\frac{a^2 \sqrt{E+P}}{c_s \mathcal{H}}=\frac{a^2 \sqrt{2XP,_X}}{c_s \mathcal{H}}
\end{equation}

\subsection{The adiabatic subtraction term for the scalar perturbation}
The Mukhanov--Sasaki equation Eq. (\ref{MSeq}) is a second derivative equation and we can derive the adiabatic subtraction term for the comoving curvature (scalar) perturbation $\mathcal{R}$ from this.
\begin{equation}
v''_k +\left(c_s^2 k^2 - \frac{z''}{z}\right)v_k=0
\tag{\ref{MSeq}}
\end{equation}
where
\begin{equation}
v_k\equiv z\mathcal{R}_k,\hspace{1em}
z\equiv\frac{a^2 \sqrt{2XP,_X}}{c_s \mathcal{H}},\hspace{1em}
c_s^2\equiv\frac{P,_X}{2XP,_{XX}+P,_X}~.
\end{equation}

In this case, using the notation in the section \ref{sec:adiabatic},
\begin{equation}
\omega_k^2=c_s^2 k^2,\hspace{1em}\sigma=-\frac{z''}{z}
\end{equation}
and the physical amplitude is schematically given by
\begin{equation}
\langle|\mathcal{R} (x)|^2\rangle_{\text{phys}} \equiv \int^{\infty}_0 \frac{dk}{2\pi^2} k^2 \left[ |\mathcal{R}_k (\eta)|^2_{\text{bare}} -|\mathcal{R}_k (\eta)|^2_{\text{sub}} \right]
\end{equation}
where
\begin{equation}
|\mathcal{R}_k (\eta)|^2_{\text{sub}} \equiv |\mathcal{R}_k (\eta)|^{2(0)} +|\mathcal{R}_k (\eta)|^{2(2)}
\label{eq:subterm}
\end{equation}
is a adiabatic solution of $\mathcal{R}$ up to second order.

Then the adiabatic subtraction term for k-inflation model can be calculated
\begin{eqnarray}
|\mathcal{R}_k (\eta)|_{\text{sub}}^2 &=&\frac{1}{2z^2}\left\{\omega_k^{-1} -\omega_k^{-2} \left(\frac{1}{2}\omega_k^{-\frac{1}{2}}\frac{d^2}{d\eta^2}\omega_k^{-\frac{1}{2}}+\frac{1}{2}\omega_k^{-1}\sigma\right)\right\}\nonumber\\
&=&\frac{1}{2z^2 c_s k}\left\{1+\frac{1}{2c_s^2 k^2}\frac{z''}{z}+\frac{1}{c_s^2 k^2}\left(\frac{1}{4}\frac{c''_s}{c_s}-\frac{3}{8}\frac{c'_s {}^2}{c_s^2}\right)\right\}~.
\label{eq:mini-sub}
\end{eqnarray}

The third term in the parenthesis in the second line comes from the derivative of $\omega_k$. In the models having canonical kinetic terms (for example slow-roll model etc.), the sound speed $c_s$ is equal to one, and this derivatives are equal to zero. This result accords with the result in \cite{Urakawa:2009-1}, which considered slow-roll inflation model.

\subsection{The adiabatic subtraction term for the tensor perturbation}
The tensor perturbation or the gravitational wave $h_{ij}$ is one of the modes of the metric perturbations Eq. (\ref{eq:metricP3}).
The tensor perturbation is gauge invariant itself, and has only two degrees of freedom because of the property of the metric.

While the scalar perturbation propagates at the sound speed, the tensor perturbation propagates at the light speed even in k-inflation case.
The each degree of freedom of tensor perturbation obeys the following equation like Mukhanov--Sasaki equation with $c_s =1$.
\begin{equation}
v''_{Tk} +\left(k^2 -\frac{a''}{a}\right)v_{Tk} =0
\end{equation}
where $v_{Tk}=\frac{a}{2} h_k$. $h_k$ denotes one of two polarized gravitational waves and we must add each results at the end of calculation.

The subtraction terms can be derived similarly.
In this cace,
\begin{equation}
\omega_k^2 =k^2,\ \sigma =-\frac{a''}{a}~.
\end{equation}
The calculation of the subtraction term has done by \cite{Urakawa:2009-1} in slow-roll model case, and the same analysis can be used in k-inflation model case.
The expression of the subtraction term for tensor perturbation is obtained as
\begin{equation}
|h_{k}(\eta)|_{\text{sub}}^2=
\frac{2}{a^2 k}\left(1+\frac{1}{2k^2}\frac{a''}{a}\right),
\end{equation}
and it is efficiently suppressed at late time.

\section{ Non-minimally coupled k-inflation}
The inflation models which has the coupling terms of inflaton and scalar curvature have been considered.
Higgs inflation model (see e.g. \cite{Bezrukov:2008-1}) is one of these models, and of course there is no reason to prohibit inflaton from coupling to scalar curvature in general.
Then let us generalize more the Lagrangian and obtain the adiabatic subtraction term for non-minimally coupled inflation models.

The action of non-minimally coupled k-inflation model is given by
\begin{equation}
S=\frac{1}{2}\int d^4 x\sqrt{-g}\left[-f(\varphi)R+2P(\varphi,X) \right]~.
\label{non-minimal-k}
\end{equation}
$f(\varphi)=1$ corresponds to the minimally coupled case and $f(\varphi)=1+\frac{1}{6}\varphi^2$ corresponds to the conformally coupled case.

In non-minimal case, the Einstein equation differs from the minimal couping case, and the scalar perturbation does not obey the Mukhanov--Sasaki equation.
However, according to \cite{Qiu:2011-1}, it is shown that the scalar curvature obeys the equation like Mukhanov--Sasaki equation by using ADM formalism \cite{Arnowitt:1959-1}.
\begin{equation}
v''_k +\left(c_{s,\text{eff}}^2 k^2 - \frac{z''_{\text{eff}}}{z_{\text{eff}}}\right) v_k =0,\ v_k \equiv z_{\text{eff}}\mathcal{R}_k
\label{eq:Qiu's}
\end{equation}
$z_{\text{eff}}$ and $c_{s,\text{eff}}$ have been estimated directly by ADM formalism in \cite{Qiu:2011-1}.
\begin{eqnarray}
z_{\text{eff}}^2 &=& 6e^{2\theta}\left(\frac{\mathcal{H}}{\theta'}-1\right)^2+2\frac{a^4 \Sigma}{\theta'^2}\label{eq:zeff}\\
c_{s,\text{eff}}^2 &=& \frac{\theta'^2-\theta''}{3(\mathcal{H}-\theta')^2+\frac{a^4 \Sigma}{e^{2\theta}}}
\label{eq:ceff}
\end{eqnarray}
where
\begin{equation}
\theta\equiv\frac{1}{2}\ln{(fa^2)},\hspace{1em}
\Sigma\equiv XP,_X +2X^2 P,_{XX}~.
\label{eq:thetasigma}
\end{equation}
i.e., $\theta'$ / $\dot{\theta}$ corresponds to $\mathcal{H}$ / $H$ respectively in the minimal coupling case.

Using Eq. (\ref{eq:Qiu's}), we can get the subtraction terms for non-minimal coupled case,
\begin{equation}
|\mathcal{R}_k (\eta)|_{\text{sub}}^2=
\frac{1}{2z_{\text{eff}}^2 c_{s,\text{eff}} k}\left\{1+\frac{1}{2c_{s,\text{eff}}^2 k^2}\frac{z''_{\text{eff}}}{z_{\text{eff}}}+\frac{1}{c_{s,\text{eff}}^2 k^2}\left(\frac{1}{4}\frac{c''_{s,\text{eff}}}{c_{s,\text{eff}}}-\frac{3}{8}\frac{c'_{s,\text{eff}} {}^2}{c_{s,\text{eff}}^2}\right)\right\}~.
\label{sub:jordan}
\end{equation}

This subtraction term depends on time and $f(\varphi)$ in a complex manner.
Moreover, the ADM formalism uses a very long method of calculation to derive the equations of motion.
Once we have the expression Eq. (\ref{sub:jordan}), there is no need to search for other methods.
However, for checking physicality and simplifying the process, we will consider a different method to obtain the subtraction terms for non-minimal coupling case in Part 3 (based upon using the equations of motion).
\section{ Time development of the subtraction terms\label{sec:timedev}}
\subsection{Freezing out and the subtraction terms}

The bare power spectrum has a property whereby certain parameters ``freeze out'': they do not develop after horizon crossing $aH=k$ in canonical case, or the sound horizon crossing $aH=c_s k$ in the non-canonical case. The reason is that the comoving curvature perturbation $\mathcal{R}$ becomes constant in the large scale limit $-k\eta \ll 1$ or $-c_s k\eta \ll 1$.

\begin{figure}[h] \begin{center}
    \includegraphics[width=7cm]{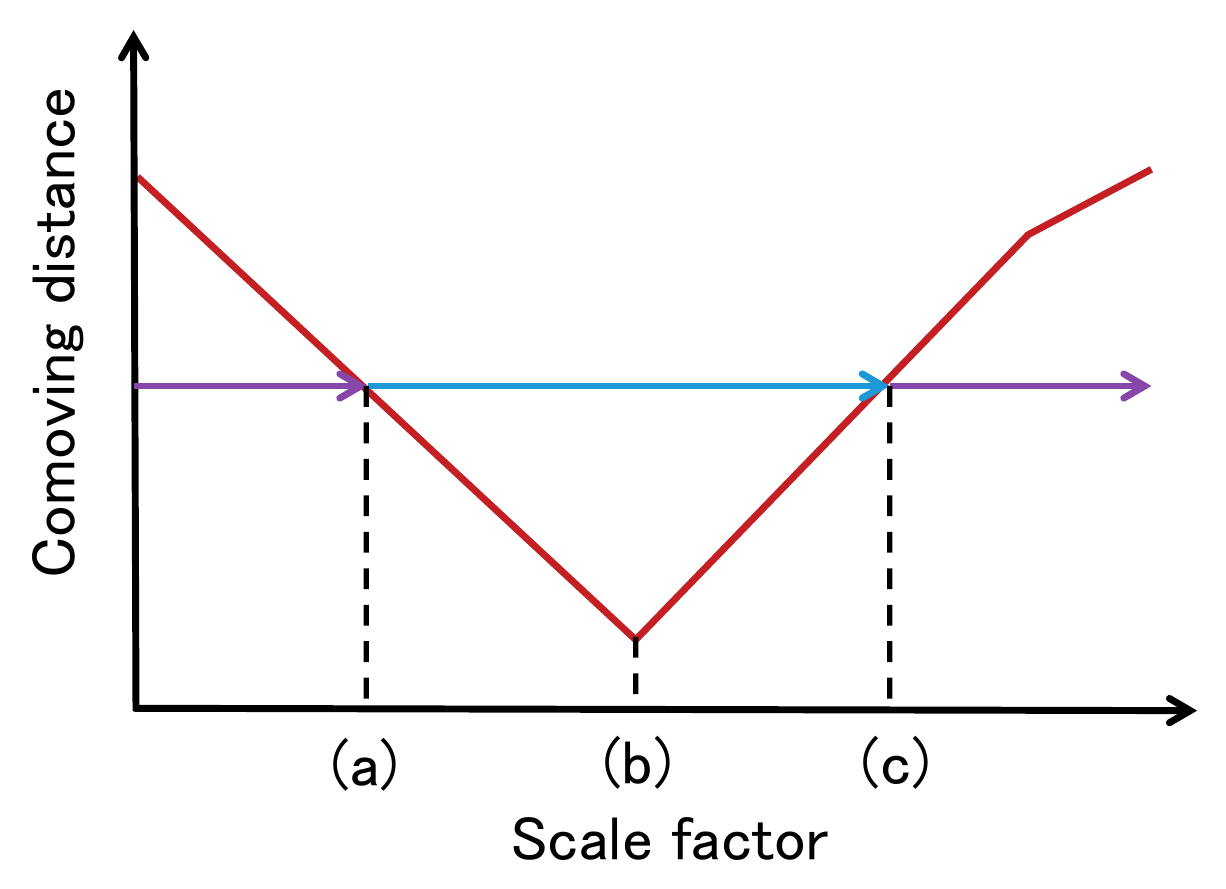}
    \caption{The time development of the comoving Hubble radius $1/aH$ (log-log graph).}
    \label{fig:crossing}
\end{center} \end{figure}

The frozen modes of the perturbation become observable after the horizon re-entry, which occurs in radiation dominant era or matter dominant era, see Fig. (\ref{fig:crossing}).
The arrows indicate a specific comoving scale $1/k$ of the power spectrum. 
The comoving horizon (or the comoving Hubble radius) shrinks during the inflation and the scale exits to the outside of the horizon at (a).
After reheating around (b), the radiation dominant era begin and the comoving Hubble radius grows.
The mode re-enter into the horizon at (c).
It is considered that the primordial perturbations are transformed into the observables with some processes after (a). 
However, it is unclear when it occurred.

The longer the time the perturbations are freezing, the more primordial information we can obtain.
Therefore we are interested in the large scale limit.

When the adiabatic regularization scheme was adopted to evaluate the power spectrum, it was considered that the subtraction terms also freeze out after crossing at first \cite{Parker:2007-1}.
The subtraction terms for slow-roll inflation models at horizon crossing are not small and cannot be neglected, therefore the author claimed, see \cite{Parker:2007-1}, that the amplitude of the perturbations was changed from the bare one.

However, the time dependence of the subtraction term is not obvious from Eq. (\ref{eq:mini-sub}) even if we take the large scale limit.
Moreover, it should be estimated up to the stage of reheating or up to times just before the primordial perturbations are transformed into classical quantities.

The method of subtraction has been argued by various authors (see the review \cite{BasteroGil:2013-1}). Based on the above, we adopt the scheme adopted by Urakawa and Starobinsky \cite{Urakawa:2009-1}.
Their scheme is similar to Parker's one, but not quite the same. In this scheme:
\begin{itemize}
\item We subtract up to the second adiabatic order term for the power spectra.
\item We consider the time development of the subtraction terms even after horizon crossing.
\item To consider time development, we do not require any cutoff of the range of the $k$ integrals.
\end{itemize}
The first condition is the minimal subtraction scheme as usual.
The second condition is different from the original idea of Parker \cite{Parker:2007-1}, while the third condition is required to maintain the equation of motion of the inflaton in coordinate space \cite{BasteroGil:2013-1}.

Importantly, we are of the view that this scheme is simple and does not need any fictitious assumptions or processes; hence we use this method to analyze the observable power spectrum.

In the following sections, we see how large a subtraction term remains after the inflation era.
Although it may be not so important for observational physics, we also investigate the time development of the subtraction term during inflation in two specific cases, see Appendix \ref{chap:sub-dev}.

\subsection{The subtraction term in terms of the slow-roll parameters}
It is useful to rewrite the subtraction term in terms of the slow-roll parameters which we are familiar with to see the time development.
There are several ways to define the slow-roll parameters.
Among them, we use the Hubble flow functions and the sound flow functions.
The Hubble flow functions $\epsilon_i$ is defined by
\begin{equation}
\epsilon_{n+1}\equiv\frac{d\ln{\epsilon_n}}{dN},\hspace{1em}\epsilon_0 = \frac{H_{\text{ini}}}{H}
\end{equation}
and the sound flow functions $\delta_i$ is defined by
\begin{equation}
\delta_{n+1}\equiv\frac{d\ln{\delta_n}}{dN},\hspace{1em}\delta_0 = \frac{c_{s,\text{ini}}}{c_s}
\label{eq:soundflow-def}
\end{equation}
where $N=\ln{a/a_{\text{ini}}}$ is $e$-folding number.
The subscript ${}_{\text{ini}}$ means the initial value which is irrelevant, but we should fix it at the horizon exit to compare the subtraction term to the frozen bare spectrum.

The definitions are also written in alternative forms
\begin{equation}
\epsilon_{n+1}=\frac{1}{H}\frac{\dot{\epsilon_n}}{\epsilon_n}=\frac{1}{\mathcal{H}}\frac{\epsilon'_n}{\epsilon_n},\hspace{1em}
\delta_{n+1}=\frac{1}{H}\frac{\dot{\delta_n}}{\delta_n}=\frac{1}{\mathcal{H}}\frac{\delta'_n}{\delta_n}~.
\end{equation}

Then rewrite $z$ in terms of $\epsilon_i$. First,
\begin{equation}
\epsilon_1 =\frac{1}{H}\frac{\dot{\epsilon_0}}{\epsilon_0}
=-\frac{\dot{H}}{H^2}~.
\end{equation}
This is equal to the slow-roll parameter $\varepsilon\equiv-\frac{\dot{H}}{H^2}$ defined in usual.

From Friedmann equations Eq. (\ref{eq:Friedmann1}) and Eq. (\ref{eq:Friedmann2})\footnote{In first order perturbation theory, it is enough considering only the zeroth order part of $z$ because the comoving scalar curvature $\mathcal{R}$ itself is the first order of perturbation.},
\begin{equation}
\dot{E}=-3H(E+P)=6H\dot{H}
\end{equation}
\begin{equation}
\therefore E+P=-2\dot{H}
\end{equation}
Substitute this into Eq. (\ref{eq:zdef}),
\begin{equation}
z\equiv\frac{a^2 \sqrt{E+P}}{c_s \mathcal{H}}=\frac{a \sqrt{E+P}}{c_s H}
=\frac{a}{c_s}\sqrt{\frac{-2\dot{H}}{H^2}}=\frac{\sqrt{2\epsilon_1}a}{c_s}
\label{eq:z-epsilon}
\end{equation}

Using this expression, we can rewrite $z'' /z$ \cite{Lorenz:2008-1}.
\begin{equation}
\frac{z''}{z}=a^2 H^2 \left\{2-\epsilon_1 +\frac{3}{2}\epsilon_2 +\frac{1}{4}\epsilon_2^2 -\frac{1}{2}\epsilon_1\epsilon_2 +\frac{1}{2}\epsilon_2 \epsilon_3 +(3-\epsilon_1 +\epsilon_2)\delta_1 +\delta^2_1 +\delta_1 \delta_2 \right\}
\label{eq:explicit-effp}
\end{equation}

We can also rewrite the third them of Eq. (\ref{eq:mini-sub}).
\begin{equation}
\frac{1}{4}\frac{c''_s}{c_s}-\frac{3}{8}\frac{c'_s {}^2}{c_s^2}
=-\frac{1}{8}a^2 H^2\delta_1 (2-2\epsilon_1+\delta_1+2\delta_2)
\label{eq:explicit-third}
\end{equation}
The calculation in detail is given Appendix \ref{app:detail} for completeness.

Finally, we obtain the expression of the adiabatic subtraction term.
\begin{equation}
|\mathcal{R}_k (\eta)|_{\text{sub}}^2
=\frac{1}{2z^2 c_s k}\left\{1+\left(\frac{aH}{c_s k}\right)^2 \left(1+\delta\epsilon+\delta c_s \right)\right\}
\label{eq:sub-in-sr}
\end{equation}
Here we defined two variables:
\begin{equation}
\delta\epsilon \equiv \frac{1}{2}\left(-\epsilon_1 +\frac{3}{2}\epsilon_2 +\frac{1}{4}\epsilon_2^2 -\frac{1}{2}\epsilon_1\epsilon_2 +\frac{1}{2}\epsilon_2 \epsilon_3\right)
\label{eq:del-e}
\end{equation}
\begin{equation}
\delta c_s \equiv
\frac{1}{8}\delta_1 (10-2\epsilon_1 +4\epsilon_2 +3\delta_1 +2\delta_2)
\label{eq:del-c}
\end{equation}
and have isolated the explicit effect of the non-canonical kinetic terms.

 During inflation, we usually expect small slow-roll parameters, which can be constructed by the Hubble flow functions and the sound flow functions, and $\delta\epsilon$ and $\delta c_s$ are smaller than $\mathcal{O}(1)$.
Even after inflation, $\delta\epsilon$ is not so large because $\epsilon_1 =2$ and $\epsilon_2 =0$ in the radiation dominant era $a(t)\propto\sqrt{t}$, and $\epsilon_1 =3/2$ and $\epsilon_2 =0$ in the matter dominant era $a(t)\propto t^{2/3}$.
How large $\delta c_s$ is depends on the model.
However, the ``sound speed'' $c_s$ does not exceed $1$ in usual k-inflation models, and we expect that $\delta c_s$ does not become too large.

Therefore it depends on the behavior of the factors $\frac{1}{2z^2 c_s k}$ and $\frac{1}{2z^2 c_s k}\left(\frac{aH}{c_s k}\right)^2$ how the subtraction term develops.
Let us see how the each term behaves at late time.

The zeroth order subtraction term $\frac{1}{2z^2 c_s k}$ becomes exponentially small in general because
\begin{equation}
\frac{1}{2z^2 c_s k}=\frac{c_s}{4\epsilon_1 a^2 k}
\end{equation}
and $a$ expands exponentially during inflation.
$\epsilon_1$ also become large 
while $c_s$ cannot become too large.
This is a natural consequence because the zeroth order term corresponds to the zero-point energy in the Minkowski spacetime.

Concerning the second order subtraction term, mainly the behavior depends on $c_s$ because it is known that $\frac{1}{2z^2 k}\left(\frac{aH}{k}\right)^2=\frac{H^2}{4\epsilon_1 k^3}$ at the late time is sufficiently suppressed compared to the its value at the horizon crossing in canonical case \cite{Urakawa:2009-1}.
Because $\epsilon_1$ in post-inflation era is bigger than during inflation and $H\propto t^{-1}$ in radiation and matter dominant era.

In the non-canonical case,
\begin{equation}
\frac{1}{2z^2 c_s k}\left(\frac{aH}{c_s k}\right)^2
=\frac{H^2}{4\epsilon_1 c_s k^3}~.
\label{eq:second-factor}
\end{equation}
Therefore if there is a model in which $c_s$ becomes sufficiently small, the subtraction terms of the model can be large enough such that it affects the power spectrum.

\newpage\part{Conformal transformations and physicality}
\setcounter{section}{0}

In Part 2, we discussed how the adiabatic subtraction term for the non-minimal k-inflation model can be derived by using the ADM formalism. 
However, minimal coupling models are studied more often than non-minimal inflation models, where part of the reason is the calculations are easier. 
Therefore, it is convenient to analyze non-minimal models if we can apply knowledge from minimal inflation models in some form.

Using a conformal transformation, we can rewrite the Lagrangian of the non-minimal coupling model as that of one in the minimal coupling model.
The metric and the pressure are altered by such a conformal transformation, hence the physical properties of the conformal-transformed universe are 
different from the true minimal coupling case, even though the action has the same form as a minimal coupling model.

The frame which has the non-minimal coupling term is denoted the Jordan, while the conformally-transformed frame is denoted the Einstein frame. It is interesting to note that the physical interpretation is completely altered, yet the observations do not change in the transformation (see e.g. \cite{Domenech:2015-1}).

Importantly, in this chapter, we will show that we can derive the adiabatic subtraction terms (extracted from the Mukhanov--Sasaki equation) for non-minimal k-inflation with the conformal transformation, without using the ADM formalism.

\section{ Einstein equation for non-minimal coupling model in Jordan frame}
Before the discussion of the conformal transformation, let us review the calculations in Jordan frame.
How different it is from the minimal case is also argued.

The action in the Jordan frame is
\begin{equation}
S=\frac{1}{2}\int d^4 x \sqrt{-g} [-f(\varphi)R+2P(\varphi,X)]
\tag{\ref{non-minimal-k}}
\end{equation}
and the Einstein equation becomes
\begin{equation}
g_{\mu\nu}\nabla_{\lambda}\nabla^{\lambda}f-\nabla_{\mu}\nabla_{\nu}f
+f\left(R_{\mu\nu}-\frac{1}{2}Rg_{\mu\nu}\right)
-P,_{X}\nabla_{\mu}\varphi\nabla_{\nu}\varphi+Pg_{\mu\nu}=0~.
\label{eq:einnon}
\end{equation}
To simplify, define three tensors and rewrite the Einstein equation as follows:
\begin{eqnarray}
G_{\mu\nu}&\equiv&R_{\mu\nu}-\frac{1}{2}Rg_{\mu\nu}\label{eq:g}\\
T_{\mu\nu}&\equiv&P,_{X}\nabla_{\mu}\varphi\nabla_{\nu}\varphi-Pg_{\mu\nu}\label{eq:p}\\
F_{\mu\nu}&\equiv&g_{\mu\nu}\nabla_{\lambda}\nabla^{\lambda}f-\nabla_{\mu}\nabla_{\nu}f\label{eq:f}
\end{eqnarray}
\begin{equation}
fG^{\mu}_{\ \nu}=T^{\mu}_{\ \nu}-F^{\mu}_{\ \nu}
\label{eq:EinsteinT}
\end{equation}

If we pick up only the first order part of the perturbations,
\begin{equation}
\delta (fG^{\mu}_{\ \nu})=G^{\ \mu}_{c\ \nu}\delta f+f_c\delta G^{\mu}_{\ \nu}
=\delta T^{\mu}_{\ \nu}-\delta F^{\mu}_{\ \nu}~.
\end{equation}
Hereafter the subscript $c$ denotes the classical part and $\delta$ denotes the first order part of the perturbations respectively.

The gauge fixing is needed to calculate the equation of the scalar perturbation.
We use the conformal-Newtonian gauge Eq. (\ref{eq:CNgauge}) in this section because the correspondence to gauge-invariant variables is clearer than other gauges.

From the transformation law of the tensor, the gauge-invariant $\overline{\delta G}^{\mu}_{~\nu}$, $\overline{\delta T}^{\mu}_{~\nu}$, and $\overline{\delta F}^{\mu}_{~\nu}$ is written by
\begin{eqnarray}
\overline{\delta G}^0_{\ 0} &\equiv& \delta G^0_{\ 0} -(G^{\ 0}_{c\ 0})' (B-\tilde{E}')\nonumber\\
\overline{\delta G}^i_{\ j} &\equiv& \delta G^i_{\ j} -(G^{\ i}_{c\ j})' (B-\tilde{E}') \label{eq:gtensor}\\
\overline{\delta G}^0_{\ i} &\equiv& \delta G^0_{\ i} -\left(G^{\ 0}_{c\ 0}-\frac{1}{3}G^{\ k}_{c\ k}\right) (B-\tilde{E}')\nonumber
\end{eqnarray}
and so on.
The gauge invariant $\overline{\delta f}$ is also obtained from the transformation law of the scalar,
\begin{equation}
\overline{\delta f}=\delta f-f'_c (B-\tilde{E})~.
\end{equation}

In conformal-Newtonian gauge, $B=\tilde{E}=0$, and we get the gauge-invariant Einstein equation by replacing the tensors by the gauge-invariant ones.
\begin{equation}
G^{\ \mu}_{c\ \nu}\overline{\delta f}+f_c\overline{\delta G}^{\mu}_{\ \nu}
=\overline{\delta T}^{\mu}_{\ \nu}-\overline{\delta F}^{\mu}_{\ \nu}~.
\end{equation}

Then calculate the tensors and the equation in accordance with the definitions Eq. (\ref{eq:g})--(\ref{eq:EinsteinT}).
After the tedious calculations, we can obtain $\overline{\delta G}^{\mu}_{\ \nu}$ \cite{Mukhanov200511}
\begin{eqnarray}
\overline{\delta G}^0_{\ 0} &=& \frac{2}{a^2} \{\Delta\Psi-3\mathcal{H}(\Psi'+\mathcal{H}\Phi)\}\\
\overline{\delta G}^0_{\ i} &=& \frac{2}{a^2}(\Psi'+\mathcal{H}\Phi),_{i}\\
\overline{\delta G}^i_{\ j} &=& -\frac{2}{a^2}\left[\left\{\Psi''+\mathcal{H}(2\Psi+\Phi)'+(2\mathcal{H}'+\mathcal{H}^2)\Phi+\frac{\Delta (\Phi-\Psi)}{2}\right\}\delta_{ij}\right.\nonumber\\
&&\left.-\frac{(\Phi-\Psi),_{ij}}{2}\right]
\label{eq:Gpertss}
\end{eqnarray}
where $\Delta\equiv\partial_{\mu}\partial^{\mu}$ and $\overline{\delta T}^{\mu}_{\ \nu}$.
\begin{eqnarray}
\overline{\delta T}^0_{\ 0} &=& 2X\delta P,_X -\delta P\\
\overline{\delta T}^0_{\ i} &=& \frac{1}{a^2}P_{c,X}\varphi'\delta\varphi,_i\\
\overline{\delta T}^i_{\ j} &=& -\delta P \delta^i_{\ j}
\end{eqnarray}
The Einstein equations are then as follows:
\begin{eqnarray}
\frac{3}{2}\mathcal{H}^2 \overline{\delta f} +
f_c\{\Delta\Psi-3\mathcal{H}(\Psi'+\mathcal{H}\Phi)\}
&=&\frac{1}{2}a^2 (\overline{\delta T}^{0}_{\ 0}-\overline{\delta F}^{0}_{\ 0})\\
f_c (\Psi'+\mathcal{H}\Phi),_i &=&\frac{1}{2}a^2 (\overline{\delta T}^{0}_{\ i}-\overline{\delta F}^{0}_{\ i})
\end{eqnarray}
\begin{eqnarray}
\left(\mathcal{H}'+\frac{1}{2}\mathcal{H}^2\right)\overline{\delta f}\delta_{ij}
&+&f_c\left\{\Psi''+\mathcal{H}(2\Psi+\Phi)'+(2\mathcal{H}'+\mathcal{H}^2)\Phi+\frac{1}{2}\Delta (\Phi-\Psi)\right\}\delta_{ij}\nonumber \\
&-&\frac{1}{2}f_c(\Phi-\Psi),_{ij}
=-\frac{1}{2}a^2 (\overline{\delta T}^{i}_{\ j}-\overline{\delta F}^{i}_{\ j})
\label{eq:third}
\end{eqnarray}

Let us set $i\neq j$ in the Eq. (\ref{eq:third}).
Because the non-diagonal content of $\overline{\delta T}^i_{~j}$ is zero by definition then
\begin{equation}
f_c(\Phi-\Psi),_{ij}=-a^2 \overline{\delta F}^{i}_{\ j}~.
\label{eq:third'}
\end{equation}
The spatial non-diagonal components of the Einstein equations are zero in the minimal coupling case.
The existence of this anisotropic inertia is a big difference between the minimal and non-minimal couplings model.

The explicit form of $F^{\mu}_{\ \nu}$ depends on the model, but we calculate the general form for generality.

From Eq. (\ref{eq:f}),
\begin{eqnarray}
F^{\mu}_{~\nu} &=&\delta^{\mu}_{~\nu}g_c^{\lambda\rho}\left(\partial_{\rho}\partial_{\lambda}-^{(0)}\!\Gamma^{\alpha}_{\rho\lambda}\partial_{\alpha}\right)f_c-g_c^{\mu\rho}\left(\partial_{\rho}\partial_{\nu}-^{(0)}\!\Gamma^{\alpha}_{\rho\nu}\partial_{\alpha}\right)f_c\nonumber\\
    &+&\delta^{\mu}_{~\nu}\left(g_c^{\lambda\rho}\partial_{\rho}\partial_{\lambda}\delta f
    +\delta g^{\lambda\rho}\partial_{\rho}\partial_{\lambda}f_c
    -g_c^{\lambda\rho} {}^{(0)}\Gamma^{\alpha}_{\rho\lambda}\partial_{\alpha}\delta f
    -g_c^{\lambda\rho}\delta\Gamma^{\alpha}_{\rho\lambda}\partial_{\alpha} f_c
    -\delta g^{\lambda\rho} {}^{(0)}\Gamma^{\alpha}_{\rho\lambda}\partial_{\alpha} f_c
\right)\nonumber\\
    &-&\!\!\!g_c^{\mu\rho}\partial_{\rho}\partial_{\nu}\delta f
    -\delta g^{\mu\rho}\partial_{\rho}\partial_{\nu}f_c
    +g_c^{\mu\rho} {}^{(0)}\Gamma^{\alpha}_{\rho\nu}\partial_{\alpha}\delta f
    +g_c^{\mu\rho}\delta\Gamma^{\alpha}_{\rho\nu}\partial_{\alpha} f_c
    +\delta g^{\mu\rho} {}^{(0)}\Gamma^{\alpha}_{\rho\nu}\partial_{\alpha} f_c
\end{eqnarray}
where the Christoffel symbol $\Gamma^{\alpha}_{\rho\lambda}=^{(0)}\!\!\Gamma^{\alpha}_{\rho\lambda}+\delta\Gamma^{\alpha}_{\rho\lambda}$ (the notation 
is temporarily changed for simplicity).
The first line is the classical part $F^{\ \mu}_{c\ \nu}$, and the rest is the linear order part $\delta F^{\mu}_{\ \nu}$.
In up to first order perturbation theory, $\overline{\delta F^{\mu}_{\ \nu} (\delta f)}=\delta F^{\mu}_{\ \nu}(\overline{\delta f})$ and we get
\begin{eqnarray}
\overline{\delta F}^{0}_{~0}&=&\frac{1}{a^2}\left(-\Delta\overline{\delta f}+3\mathcal{H}\overline{\delta f}'-6\mathcal{H}\Phi f'_c-3\Psi'f'_c\right)\\
\overline{\delta F}^{0}_{~i}&=&\frac{1}{a^2}\left(-\overline{\delta f}'+\mathcal{H}\overline{\delta f}+\Phi f'_c\right),_i\\
\overline{\delta F}^{i}_{~j}&=&\frac{1}{a^2}\left[\delta^i_{~j}\left\{\overline{\delta f}''-\Delta\overline{\delta f}-2\Phi f''_c+\mathcal{H}\overline{\delta f}'-(\Phi+2\Psi)'f'_c-2\mathcal{H}\Phi f'_c\right\}\right.\nonumber\\
&&\left.+\partial_i \partial_j \overline{\delta f}\right]
\end{eqnarray}

Then when $i\neq j$,
\begin{equation}
f_c(\Phi-\Psi),_{ij}=-a^2 \overline{\delta F}^{i}_{\ j}=-\partial_i \partial_j \overline{\delta f}
\label{eq:third-3}
\end{equation}
In the minimal coupling case, there is no anisotropic inertia and we can combine the Einstein equations to get the Mukhanov--Sasaki 
equation \cite{Mukhanov200511}. However, due to the anisotropic inertia term and other additional terms from $\delta f$, it is difficult to 
combine straightforwardly the equations in the non-minimal coupling case without resorting to different mathematical techniques, such as 
the ADM formalism.

\section{ Conformal transformation and invariance}
\subsection{Conformal transformation}
Again consider the non-minimal coupling k-inflation model in the Jordan frame.
\begin{equation}
S=\frac{1}{2}\int d^4 x\sqrt{-g}\left[-f(\varphi)R+2P(\varphi,X) \right]
\tag{\ref{non-minimal-k}}
\end{equation}
To derive the adiabatic subtraction term without ADM formalism, perform the conformal transformation 
$g_{\mu\nu}\rightarrow \widehat{g}_{\mu\nu}=f(\varphi)g_{\mu\nu}$ to obtain the action in the Einstein frame \cite{Wald198406,Kubota:2012-1}:
\begin{equation}
S=\frac{1}{2}\int d^4 \widehat{x}\sqrt{-\widehat{g}}\left[-\widehat{R}+2\widehat{P}(\varphi,\widehat{X}) \right]
\label{eq:Eaction}
\end{equation}
where
\begin{eqnarray}
\widehat{a}(\widehat{\eta})&\equiv& \sqrt{f}a(\eta)\nonumber\\
\widehat{R}&\equiv&\frac{1}{f}\left[R-3\nabla_{\mu}\nabla^{\mu}\ln{f}-\frac{3}{2}(\partial_{\mu}\ln{f})(\partial^{\mu}\ln{f})\right]\nonumber\\
\widehat{X}&\equiv&\frac{1}{2f}g^{\mu\nu}\partial_{\mu}\varphi\partial_{\nu}\varphi=\frac{1}{f}X\nonumber\\
\widehat{P}&\equiv&\frac{1}{f^2}P+\frac{3}{4f}(\partial_{\mu}\ln{f})(\partial^{\mu}\ln{f})~.
\label{eq:translaw}
\end{eqnarray}

Note that the coordinates are not changed by the conformal transformation when we use conformal time.
i.e.) $d\widehat{\eta}=d\eta$, $d\widehat{x}^i =dx^i$.
Then the line element in the conformal-Newtonian gauge becomes
\begin{eqnarray}
d\widehat{s}^2 &\equiv& f(\varphi)ds^2\nonumber\\
&=& f(\varphi)a^2 (\eta)\left[(1+2\Phi)d\eta^2 -(1-2\Psi)d\vec{x}^2\right]\nonumber\\
&=& \widehat{a}^2 (\eta)\left[(1+2\widehat{\Phi})d\eta^2 -(1-2\widehat{\Psi})d\vec{x}^2\right]~.
\end{eqnarray}
After expanding each coefficient up to linear order of perturbations, we can get the transformation laws of the scalar perturbations \cite{Chiba:2008-1}.
\begin{equation}
\widehat{\Phi}=\Phi+\frac{\overline{\delta f}}{2f_c},~
\widehat{\Psi}=\Psi-\frac{\overline{\delta f}}{2f_c}
\label{eq:transPhiPsi}
\end{equation}

Let us check whether or not the anisotropic inertia in the Jordan frame vanishes in the Einstein frame.
Substituting Eq. (\ref{eq:transPhiPsi}) into Eq. (\ref{eq:third-3}) leads to
\begin{equation}
f_c\left[\left(\widehat{\Phi}-\frac{\overline{\delta f}}{2f_c}\right)-\left(\widehat{\Psi}+\frac{\overline{\delta f}}{2f_c}\right)\right],_{ij}=f_c\left(\widehat{\Phi}-\widehat{\Psi}\right),_{ij}-\partial_i \partial_j \overline{\delta f}=-\partial_i \partial_j \overline{\delta f}~.
\end{equation}
Therefore, we can find the correct relation: $\widehat{\Phi}=\widehat{\Psi}$ in the Einstein frame.

\subsection{Conformal invariance and physicality\label{sec:conformal}}

The conformal invariance of the comoving curvature perturbation $\mathcal{R}$ and its correlation functions 
$\langle\mathcal{R}\dots\rangle$ was shown in \cite{Chiba:2008-1,Kubota:2012-1}. i.e.,
\begin{equation}
\widehat{\mathcal{R}}=\mathcal{R},\hspace{1em}
\langle\widehat{\mathcal{R}}\dots\rangle = \langle\mathcal{R}\dots\rangle
\end{equation}
Then, if we regularize the correlation functions by adiabatic regularization, the subtraction terms in each frame must also agree with each other. Because otherwise the regularized power spectra depends on the frames and the method of either conformal transformation or the adiabatic regularization is spoiled.

In this section, we will derive the adiabatic subtraction terms in the Einstein frame and show the conformal invariance of the 
adiabatic subtraction terms in both the Jordan and Einstein frames.

First, consider the equation from the action Eq. (\ref{eq:Eaction})
\begin{equation}
\frac{d^2}{d\widehat{\eta}^2}\widehat{v}_k +\left(\widehat{c}_s^2 k^2 - \frac{1}{\widehat{z}}\frac{d^2 \widehat{z}}{d\widehat{\eta}^2}\right)\widehat{v}_k=0
\end{equation}
where
\begin{equation}
\widehat{v}_k\equiv\widehat{z}\widehat{\mathcal{R}}_k =\widehat{z}\mathcal{R}_k,\hspace{1em}
\widehat{z}\equiv\frac{\sqrt[]{2\widehat{\epsilon}_1}\widehat{a}}{\widehat{c}_s},\hspace{1em}
\widehat{c}_s^2\equiv\frac{\widehat{P},_{\widehat{X}}}{2\widehat{X}\widehat{P},_{\widehat{X}\widehat{X}}+\widehat{P},_{\widehat{X}}}~.
\end{equation}
From the universality of the conformal time, the equation becomes
\begin{equation}
\widehat{v}''_k +\left(\widehat{c}_s^2 k^2 - \frac{\widehat{z}''}{\widehat{z}}\right)\widehat{v}_k=0~.
\label{eq:EMS}
\end{equation}
We can get the adiabatic subtraction terms for the power spectrum from Eq. (\ref{eq:EMS}) as the same way in the minimal coupling case.
\begin{equation}
|\widehat{\mathcal{R}} (\eta)|_{\text{sub}}^2=
\frac{1}{2\widehat{z}^2 \widehat{c}_s k}\left\{1+\frac{\widehat{z}''}{2\widehat{z}}\frac{1}{\widehat{c}_s^2 k^2}+\frac{1}{\widehat{c}_s^2 k^2}\left(\frac{1}{4}\frac{\widehat{c}''_s}{\widehat{c}_s}-\frac{3}{8}\frac{\widehat{c}'_s {}^2}{\widehat{c}_s^2}\right)\right\}
\label{sub:einstein}
\end{equation}

From the adiabatic subtraction term in the Jordan frame Eq. (\ref{sub:jordan}) and in the Einstein frame Eq. (\ref{sub:einstein}), it is shown that the adiabatic subtraction terms of non-minimally coupled k-inflation model depend on only $z_{\text{eff}}$ / $\widehat{z}$ and $c_{s,\text{eff}}$ / $\widehat{c}_s$.
Therefore, if $\widehat{z}$ and $\widehat{c}_s$ are equal to the effective quantities Eq. (\ref{eq:zeff})--(\ref{eq:ceff}) after conformal transformation, the subtraction terms in both the Jordan frame and Einstein frame correspond to each other.

Once it has shown the correspondence of the subtraction terms in both frame, it is not necessary to derive it in Jordan frame by using ADM formalism.
We can calculate it in the Einstein frame, and transform it into the Jordan frame if we want to see the physical property due to the non-minimal coupling.

For conformal transformation, it is convenient to define two functions like Eq. (\ref{eq:thetasigma}).
\begin{equation}
\theta\equiv\frac{1}{2}\ln{(fa^2)}=\frac{1}{2}\ln{(\widehat{a}^2)},\hspace{1em}
\widehat{\Sigma}\equiv\widehat{X}\widehat{P}_{,\widehat{X}}+2\widehat{X}^2 \widehat{P}_{,\widehat{X}\widehat{X}}
\end{equation}
Using Friedmann equations in the Einstein frame, $\widehat{c}_s$ and $\widehat{z}$ are expressed by
\begin{equation}
\widehat{c}_s^2 =\frac{\widehat{X}\widehat{P},_{\widehat{X}}}{\widehat{\Sigma}}=\frac{\widehat{\epsilon}_1 \widehat{H}^2}{\widehat{\Sigma}},\hspace{1em}
\widehat{z}^2 =\frac{2\widehat{\epsilon}_1 \widehat{a}^2}{\widehat{c}_s^2}=\frac{2\widehat{a}^2 \widehat{\Sigma}}{\widehat{H}^2}~.
\label{eq:Ecs}
\end{equation}
Thus we should transform three quantities $\widehat{H}$, $\widehat{\epsilon_1}$ and $\widehat{\Sigma}$.
\begin{eqnarray}
\widehat{H}&=&\frac{\mathcal{\widehat{H}}}{\widehat{a}}
    =\frac{\widehat{a}'}{\widehat{a^2}}
    =e^{-\theta}\theta'\\
\widehat{\epsilon}_1&\equiv&\frac{1}{\widehat{\mathcal{H}}}\frac{\widehat{\epsilon}'_0}{\widehat{\epsilon}_0}
    =1-\frac{\widehat{\mathcal{H}}'}{\widehat{\mathcal{H}}^2}
    =1-\frac{\theta''}{\theta' {}^{2}}\label{eq:hatep}\\
\widehat{\Sigma}&=&\frac{X}{f}\left(f\frac{\partial}{\partial X}\right)\left\{\frac{1}{f^2}P+\frac{3X}{2f}\left(\frac{d\ln{f}}{d\varphi}\right)^2\right\}\nonumber\\
&&+2\left(\frac{X}{f}\right)^2 \left(f\frac{\partial}{\partial X}\right)^2 \left\{\frac{1}{f^2}P+\frac{3X}{2f}\left(\frac{d\ln{f}}{d\varphi}\right)^2\right\}\nonumber\\
&=&\frac{1}{f^2}\left\{XP,_X+2X^2 P,_{XX}+\frac{3}{2}fX\left(\frac{d\ln{f}}{d\varphi}\right)^2\right\}\nonumber\\
&=&\frac{1}{f^2}\left\{\Sigma+\frac{3}{4f}\left(\frac{f'}{a}\right)^2\right\}\nonumber\\
&=&e^{-4\theta}a^4\left\{\Sigma+3\frac{e^{2\theta}}{a^4}(\theta'-\mathcal{H})^2\right\}\nonumber\\
&=&e^{-4\theta}\theta'^2 \left\{3e^{2\theta}\left(\frac{\mathcal{H}}{\theta'}-1\right)^2+\frac{a^4 \Sigma}{\theta'^2}\right\}
\end{eqnarray}
To transform $\widehat{\Sigma}$, we used Eq. (\ref{eq:translaw}) and the following relation.
\begin{eqnarray}
\widehat{X}&\equiv&\frac{1}{2}\widehat{g}^{\mu\nu}\partial_{\mu}\varphi\partial_{\nu}\varphi
=\frac{2}{3}\left(\frac{d\varphi}{d\ln{f}}\right)^2\cdot\frac{3}{4}\widehat{g}^{\mu\nu}(\partial_{\mu}\ln{f})(\partial_{\nu}\ln{f})\\
\widehat{P}&\equiv&\frac{1}{f^2}P+\frac{3}{4}\widehat{g}^{\mu\nu}(\partial_{\mu}\ln{f})(\partial_{\nu}\ln{f})\nonumber\\
&=&\frac{1}{f^2}P+\frac{3}{2}\widehat{X}\left(\frac{d\ln{f}}{d\varphi}\right)^2
\end{eqnarray}
Substitute them into (\ref{eq:Ecs}),
\begin{eqnarray}
\widehat{z}^2&=&2\left\{3e^{2\theta}\left(\frac{\mathcal{H}}{\theta'}-1\right)^2+\frac{a^4 \Sigma}{\theta'^2}\right\}\label{eq:zcon}\\
\widehat{c}_s^2&=&\frac{e^{2\theta}\left(1-\frac{\theta''}{\theta' {}^{2}}\right)}{3e^{2\theta}\left(\frac{\mathcal{H}}{\theta'}-1\right)^2+\frac{a^4 \Sigma}{\theta'^2}}=\frac{\theta'^2-\theta''}{3(\mathcal{H}-\theta')^2+\frac{a^4 \Sigma}{e^{2\theta}}}\label{eq:ccon}
\end{eqnarray}
Comparing the results with Eq. (\ref{eq:zeff})--(\ref{eq:ceff}), it is realized that the effective variables in the Jordan frame and the variables in the Einstein frame agree with each other.
\begin{equation}
z_{\text{eff}}=\widehat{z},\hspace{1em}c_{s,\text{eff}}=\widehat{c}_s
\end{equation}
Therefore the adiabatic subtraction terms in both Einstein and Jordan frame are identical, and we get the same results of regularized 
power spectrum in either frame.

Of course, because the comoving curvature perturbation itself is conformally invariant, the equations the perturbations obey should also be 
conformally invariant. The above calculations show this explicitly.
\newpage\part{Conclusion}
\setcounter{section}{0}

We have investigated the adiabatic regularization of the power spectrum in general single scalar field inflation.
We derived the adiabatic subtraction terms and found that (due to the sound speed) their behavior is not trivial.  
The analysis was partly done by assuming some specific models or approximations \cite{Alinea:2015-1}, and there may be the models which have growing or non-vanishing subtraction terms.

We have also calculated the subtraction terms for the non-minimally coupled inflaton case.
The subtraction term is completely determined by the equation of motion of the frame-invariant scalar curvature perturbation, and we confirmed its frame invariance explicitly.

\section*{Discussion and future work}
\subsection*{What do vanishing subtraction terms mean?}

In section \ref{sec:timedev}, we have considered the time development of the adiabatic subtraction terms.
We saw that the speed of sound may be able to make the subtraction terms large enough to affect the observable power spectrum: in canonical inflation, it generally decays after inflation era (see Appendix \ref{chap:sub-dev-after}) and we expect that it becomes sufficiently small when they are transferred to  observables.


We assumed a scheme where the bare spectrum freezes at the (sound) horizon crossing, while the subtraction terms remain time-developing.
The vanishing subtraction terms mean that the integral of the bare spectrum at the crossing has no divergence to begin with.
This result agrees with the conventional interpretation about the observable power spectrum: the observable spectrum has no high frequency modes because they have not yet exited the horizon and not yet become ``classical''.

However, even if the result with regularization does not differ from the result ``without" regularization, it does not mean that the argument for regularization of the power spectrum is unnecessary. Indeed, it gives mathematical support to the observable power spectrum, and also it is needed in an interacting theory.


\subsection*{Future work}

In section \ref{sec:timedev}, we have investigated how the subtraction terms behave qualitatively at late times with the subtraction scheme used in \cite{Urakawa:2009-1}. Related to this scheme, there are some problems still to be solved.

The subtraction scheme was not established completely.
There are many discussions about this issue (e.g. \cite{Finelli:2007-1,Urakawa:2009-1,Haro:2010-1,Marozzi:2011-1,BasteroGil:2013-1}) and some papers such as \cite{Finelli:2007-1,BasteroGil:2013-1} claim that the regularization of the power spectrum is originally unnecessary.
Some of these claims have already been rejected, nevertheless, the argument is still continuing.
This is possibly due to our poor knowledge of the ``freezing out" and the unclear nature of the renormalization condition using adiabatic subtraction (see also \cite{Markkanen:2013-1}).

Understanding  ``freezing out" is an important problem of cosmology.
While it is true that the comoving curvature perturbation loses its time-dependence at large scales: $-k\eta\ll 1$, how to become classical is still unclear.
The scheme we take in this thesis assumes that the freezing out does not occur at the horizon crossing, in a sense.
When the quantum fluctuations become observables is a fundamental problem we should consider in more detail.

We should also mention that we have only calculated the value of the subtraction terms qualitatively.
As we saw in section \ref{sec:timedev}, it is possible for terms to survive or become large when the model has non-canonical kinetic terms.
We therefore need to know how large these surviving terms are and compare the regularized power spectrum with the bare one.
Some models might need to be modified to achieve a realistic power spectrum if the subtraction terms remain significant at the time of classicalization
:
we can use the regularized power spectrum to constrain an inflationary model by combining it with other theoretical constraints.

\newpage
\appendix 
\renewcommand{\thesection}{\Alph{section}}
\renewcommand{\thesubsection}{\Alph{section}.\arabic{subsection}}
\makeatletter
    \renewcommand{\theequation}{%
    \Alph{section}.\arabic{equation}}
    \@addtoreset{equation}{section}
\makeatother
\part{Appendices}
\section{Hubble flow equations in detail}
\label{app:detail}

\subsection*{The effective potential of Mukhanov--Sasaki equation}
Let us derive the explicit form of the effective potential Eq. (\ref{eq:explicit-effp})
\begin{equation}
\frac{z''}{z}=a^2 H^2 \left\{2-\epsilon_1 +\frac{3}{2}\epsilon_2 +\frac{1}{4}\epsilon_2^2 -\frac{1}{2}\epsilon_1\epsilon_2 +\frac{1}{2}\epsilon_2 \epsilon_3 +(3-\epsilon_1 +\epsilon_2)\delta_1 +\delta^2_1 +\delta_1 \delta_2 \right\}~.
\tag{\ref{eq:explicit-effp}}
\end{equation}

From Eq. (\ref{eq:z-epsilon}),
\begin{equation}
z=\frac{\sqrt{2\epsilon_1}a}{c_s}
\tag{\ref{eq:z-epsilon}}
\end{equation}
and
\begin{equation}
z''=a\frac{d}{dt}\left(a\frac{dz}{dt}\right)=a\dot{a}\dot{z}+a^2\ddot{z}=a^2 (H\dot{z}+\ddot{z})~.
\label{eq:z-2ndd}
\end{equation}

Then let us calculate the time derivative of $z$.
After some calculation, we get
\begin{equation}
\dot{z}=z\left(H-\frac{\dot{H}}{H}+\frac{1}{2}\frac{\ddot{H}}{\dot{H}}-\frac{\dot{c_s}}{c_s}\right)~,
\end{equation}
\begin{eqnarray}
\ddot{z}&=&\dot{z}\left(H-\frac{\dot{H}}{H}+\frac{1}{2}\frac{\ddot{H}}{\dot{H}}-\frac{\dot{c_s}}{c_s}\right)+z\frac{d}{dt}\left(H-\frac{\dot{H}}{H}+\frac{1}{2}\frac{\ddot{H}}{\dot{H}}-\frac{\dot{c_s}}{c_s}\right)\nonumber\\
&=&z\left\{\left(H-\frac{\dot{H}}{H}+\frac{1}{2}\frac{\ddot{H}}{\dot{H}}-\frac{\dot{c_s}}{c_s}\right)^2 + \left(\dot{H}-\frac{\ddot{H}}{H}+\frac{\dot{H}^2}{H^2}+\frac{1}{2}\frac{H^{(3)}}{\dot{H}}-\frac{1}{2}\frac{\ddot{H}^2}{\dot{H}^2}-\frac{\ddot{c_s}}{c_s}+\frac{\dot{c_s}^2}{c_s^2}\right)\right\}\nonumber\\
&=&z\left\{H^2 -\dot{H} +\frac{H\ddot{H}}{\dot{H}}+2\frac{\dot{H}^2}{H^2}-2\frac{\ddot{H}}{H}-\frac{1}{4}\frac{\ddot{H}^2}{\dot{H}^2}+\frac{1}{2}\frac{H^{(3)}}{\dot{H}}\right.\nonumber\\
    &&\left.+\left(-2H+2\frac{\dot{H}}{H}-\frac{\ddot{H}}{\dot{H}}\right)\frac{\dot{c_s}}{c_s}+2\frac{\dot{c_s}^2}{c_s^2}-\frac{\ddot{c_s}}{c_s}\right\}~.
\end{eqnarray}

Substitute them into Eq. (\ref{eq:z-2ndd}).
\begin{eqnarray}
\frac{z''}{z}&=&\frac{a^2 (H\dot{z}+\ddot{z})}{z}=a^2 H^2 \left(\frac{\dot{z}}{Hz}+\frac{\ddot{z}}{H^2 z}\right)\nonumber\\
&=&a^2 H^2\left\{2-2\frac{\dot{H}}{H^2} +\frac{3}{2}\frac{\ddot{H}}{H\dot{H}}+2\frac{\dot{H}^2}{H^4}-2\frac{\ddot{H}}{H^3}-\frac{1}{4}\frac{\ddot{H}^2}{H^2\dot{H}^2}+\frac{1}{2}\frac{H^{(3)}}{H^2\dot{H}}\right.\nonumber\\
&&\left.+\left(-\frac{3}{H}+2\frac{\dot{H}}{H^3}-\frac{\ddot{H}}{H^2\dot{H}}\right)\frac{\dot{c_s}}{c_s}+\frac{2}{H^2}\frac{\dot{c_s}^2}{c_s^2}-\frac{1}{H^2}\frac{\ddot{c_s}}{c_s}\right\}
\label{eq:explicit-effp-t}
\end{eqnarray}

Then rewrite this in terms of the Hubble and sound flow functions.
Eq. (\ref{eq:explicit-effp-t}) contains the third derivative of $H$ and we need up to the third Hubble flow function $\epsilon_3$.

The explicit forms of the flow functions are below.
\begin{equation}
\epsilon_1 =-\frac{\dot{H}}{H^2}
\end{equation}
\begin{equation}
\epsilon_1 \epsilon_2
=\frac{\dot{\epsilon_1}}{H}
=-\frac{\ddot{H}}{H^3}+2\frac{\dot{H}^2}{H^4}
\end{equation}
\begin{equation}
\epsilon_2 = \frac{1}{H}\frac{\dot{\epsilon_1}}{\epsilon_1}=\frac{\ddot{H}}{H\dot{H}}-2\frac{\dot{H}}{H^2}
\end{equation}
\begin{equation}
\epsilon_2 \epsilon_3=\frac{\dot{\epsilon_2}}{H}=-3\frac{\ddot{H}}{H^3}+4\frac{\dot{H}^2}{H^4}-\frac{\ddot{H}^2}{H^2 \dot{H}^2}+\frac{H^{(3)}}{H^2 \dot{H}}
\end{equation}

The term $\frac{1}{2}\frac{H^{(3)}}{H^2\dot{H}}$ can be rewrote by using $\epsilon_2 \epsilon_3$, but then the term proportional to $\frac{\ddot{H}^2}{H^2\dot{H}^2}$ remains.
To rewrite it we need the square of $\epsilon_2$.
\begin{equation}
\epsilon_2^2 =\frac{\ddot{H}^2}{H^2 \dot{H}^2}+4\frac{\dot{H}^2}{H^4}-4\frac{\ddot{H}}{H^3}
\end{equation}

Rewrite the second line of Eq. (\ref{eq:explicit-effp-t}).
\begin{eqnarray}
&&2-2\frac{\dot{H}}{H^2} +\frac{3}{2}\frac{\ddot{H}}{H\dot{H}}+2\frac{\dot{H}^2}{H^4}-2\frac{\ddot{H}}{H^3}-\frac{1}{4}\frac{\ddot{H}^2}{H^2\dot{H}^2}+\frac{1}{2}\frac{H^{(3)}}{H^2\dot{H}}\nonumber\\
&=&2-2\frac{\dot{H}}{H^2} +\frac{3}{2}\frac{\ddot{H}}{H\dot{H}}-\frac{1}{2}\frac{\ddot{H}}{H^3}+\frac{1}{4}\frac{\ddot{H}^2}{H^2\dot{H}^2}+\frac{1}{2}\epsilon_2 \epsilon_3\nonumber\\
&=&2-2\frac{\dot{H}}{H^2} +\frac{3}{2}\frac{\ddot{H}}{H\dot{H}}-\frac{\dot{H}^2}{H^4}+\frac{1}{2}\frac{\ddot{H}}{H^3}+\frac{1}{4}\epsilon_2^2+\frac{1}{2}\epsilon_2 \epsilon_3\nonumber\\
&=&2-2\frac{\dot{H}}{H^2} +\frac{3}{2}\frac{\ddot{H}}{H\dot{H}}+\frac{1}{4}\epsilon_2^2-\frac{1}{2}\epsilon_1 \epsilon_2+\frac{1}{2}\epsilon_2 \epsilon_3\nonumber\\
&=&2-\epsilon_1+\frac{3}{2}\epsilon_2+\frac{1}{4}\epsilon_2^2-\frac{1}{2}\epsilon_1 \epsilon_2+\frac{1}{2}\epsilon_2 \epsilon_3
\label{eq:detail1}
\end{eqnarray}

Also up to the second sound flow function $\delta_2$ is needed.
\begin{equation}
\delta_1 =-\frac{1}{H}\frac{\dot{c_s}}{c_s}
\end{equation}
\begin{equation}
\delta_1 \delta_2
=\frac{\dot{\delta_1}}{H}
=\frac{\dot{H}}{H^3}\frac{\dot{c_s}}{c_s}-\frac{1}{H^2}\frac{\ddot{c_s}}{c_s}+\frac{1}{H^2}\frac{\dot{c_s}^2}{c_s^2}
\end{equation}

Rewrite the third line of Eq. (\ref{eq:explicit-effp-t}).
\begin{eqnarray}
&&\left(-\frac{3}{H}+2\frac{\dot{H}}{H^3}-\frac{\ddot{H}}{H^2\dot{H}}\right)\frac{\dot{c_s}}{c_s}+\frac{2}{H^2}\frac{\dot{c_s}^2}{c_s^2}-\frac{1}{H^2}\frac{\ddot{c_s}}{c_s}\nonumber\\
&=&\left(-\frac{3}{H}+\frac{\dot{H}}{H^3}-\frac{\ddot{H}}{H^2\dot{H}}\right)\frac{\dot{c_s}}{c_s}+\frac{1}{H^2}\frac{\dot{c_s}^2}{c_s^2}+\delta_1 \delta_2\nonumber\\
&=&\left(3-\frac{\dot{H}}{H^2}+\frac{\ddot{H}}{H\dot{H}}\right)\delta_1+\delta_1^2+\delta_1 \delta_2\nonumber\\
&=& (3-\epsilon_1 +\epsilon_2)\delta_1+\delta_1^2+\delta_1 \delta_2
\label{eq:detail2}
\end{eqnarray}

By combining Eq. (\ref{eq:detail1}) and eq.(\ref{eq:detail2}), we can get the explicit form of the effective potential Eq. (\ref{eq:explicit-effp}).

\subsection*{The third term of the adiabatic subtraction term}
Next, let us rewrite the third term of the subtraction term Eq. (\ref{eq:explicit-third}).
\begin{equation}
\frac{1}{4}\frac{c''_s}{c_s}-\frac{3}{8}\frac{c'_s {}^2}{c_s^2}
=- \frac{1}{8}a^2 H^2\delta_1 (2-2\epsilon_1+\delta_1+2\delta_2)
\tag{\ref{eq:explicit-third}}
\end{equation}

To use the above calculation, we need translate this terms in terms of proper time derivatives.
\begin{eqnarray}
\frac{1}{4}\frac{c''_s}{c_s}-\frac{3}{8}\frac{c'_s {}^2}{c_s^2}
&=&\frac{1}{4}\left(\frac{a\dot{a}\dot{c_s}+a^2 \ddot{c_s}}{c_s}\right)-\frac{3}{8}\frac{a^2 \dot{c_s}^2}{c_s^2}\nonumber\\
&=&a^2 H^2 \left\{\frac{1}{4}\frac{\dot{c_s}}{Hc_s}+\frac{1}{4}\frac{\ddot{c_s}}{H^2 c_s}-\frac{3}{8}\frac{\dot{c_s}^2}{H^2 c_s^2}\right\}\nonumber\\
&=&a^2 H^2 \left\{-\frac{1}{4}\delta_1 +\frac{1}{4}\delta_1 (\epsilon_1 +\delta_1 -\delta_2 )-\frac{3}{8}\delta_1^2\right\}\nonumber\\
&=&-a^2 H^2 \cdot \frac{1}{8}\delta_1 (2-2\epsilon_1+\delta_1+2\delta_2)
\label{eq:detail3}
\end{eqnarray}

Combining Eq. (\ref{eq:detail3}) with the effect by time dependent sound speed in the effective potential Eq. (\ref{eq:detail2}), we obtain
\begin{equation}
\frac{1}{2}\left\{(3-\epsilon_1 +\epsilon_2)\delta_1+\delta_1^2+\delta_1 \delta_2\right\}
-\frac{1}{8}\delta_1 (2-2\epsilon_1+\delta_1+2\delta_2)
=\frac{1}{8}\delta_1 (10-2\epsilon_1 +4\epsilon_2 +3\delta_1 +2\delta_2)~.
\end{equation}

Then we have rewritten the adiabatic subtraction term in terms of the flow functions
\begin{equation}
|\mathcal{R}_k (\eta)|_{\text{sub}}^2
=\frac{1}{2z^2 c_s k}\left\{1+\left(\frac{aH}{c_s k}\right)^2 \left(1+\delta\epsilon+\delta c_s \right)\right\}
\tag{\ref{eq:sub-in-sr}}
\end{equation}
where
\begin{equation}
\delta\epsilon \equiv \frac{1}{2}\left(-\epsilon_1 +\frac{3}{2}\epsilon_2 +\frac{1}{4}\epsilon_2^2 -\frac{1}{2}\epsilon_1\epsilon_2 +\frac{1}{2}\epsilon_2 \epsilon_3\right)~,
\tag{\ref{eq:del-e}}
\end{equation}
\begin{equation}
\delta c_s \equiv
\frac{1}{8}\delta_1 (10-2\epsilon_1 +4\epsilon_2 +3\delta_1 +2\delta_2)~.
\tag{\ref{eq:del-c}}
\end{equation}
\newpage\section{The time development of the adiabatic subtraction terms during inflation}\label{chap:sub-dev}
We have investigated the time dependence of the adiabatic subtraction term for non-canonical inflation in the section \ref{sec:timedev}.
The zeroth adiabatic order term decays in general, while we cannot say anything about the second order term unless we specify the models (see \cite{Alinea:2015-1}).

From the view point we take in the section \ref{sec:timedev}, the time dependence of the subtraction term during inflation is irrelevant.
However, some people still claim that the subtraction term should be estimated at the horizon exit \cite{Agullo:2010-1,Agullo:2011-1}.

Hence we estimate the time development of the subtraction term during inflation in the following two cases: (1) the slow-roll inflation model in quasi-de Sitter spacetime and (2) DBI inflation model.
As a result, we find that the subtraction terms at the horizon exit exceedingly suppress the bare power spectrum in these two cases.

\subsection{The slow-roll inflation model in quasi-de Sitter spacetime\label{sec:quasi-dS}} 

Let us review the case of slow-roll inflation model in de Sitter background spacetime first.
Of course, however, the exact de Sitter universe can be achieved by the cosmological constant rather than inflaton.
We set the slow-roll inflaton with perturbations on a de Sitter background, and hence the derivation of the power spectrum is an approximation.

In the case of exact de Sitter universe,
\begin{equation}
a(t)= e^{Ht} \longrightarrow a(\eta)=-\frac{1}{H\eta}
\label{eq:exactdS}
\end{equation}
and $H$ is constant.
The Hubble flow functions $\epsilon_i$ ($i\geq 1$) are equal to zero by definition.
The sound flow functions are also zero.
Furthermore, we are assuming a model with canonical kinetic term.

Then the effective potential of Mukhanov--Sasaki equation becomes
\begin{equation}
\frac{z''}{z}=2a^2 H^2=\frac{2}{\eta^2}
\end{equation}
and the following expression is obtained.
\begin{equation}
|\mathcal{R}_k (\eta)|_{\text{sub}}^2=\frac{H^2}{2a^2\dot{\varphi}^2 k}\left(1+\frac{1}{k^2 \eta^2}\right)
=\frac{H^4}{2\dot{\varphi}^2 k^3}\left(k^2 \eta^2 +1\right)
\label{eq:sub-dS}
\end{equation}
Here we use that $E+P=\dot{\varphi}^2$.
In the large scale limit,
\begin{equation}
\lim_{-k\eta \to 0}|\mathcal{R}_k (\eta)|_{\text{sub}}^2
=\frac{H^4}{2\dot{\varphi}^2 k^3}~.
\end{equation}
This is equal to the bare power spectrum of the slow-roll model in the large scale limit.

\begin{figure}[h] \begin{center}
    \includegraphics[width=10cm]{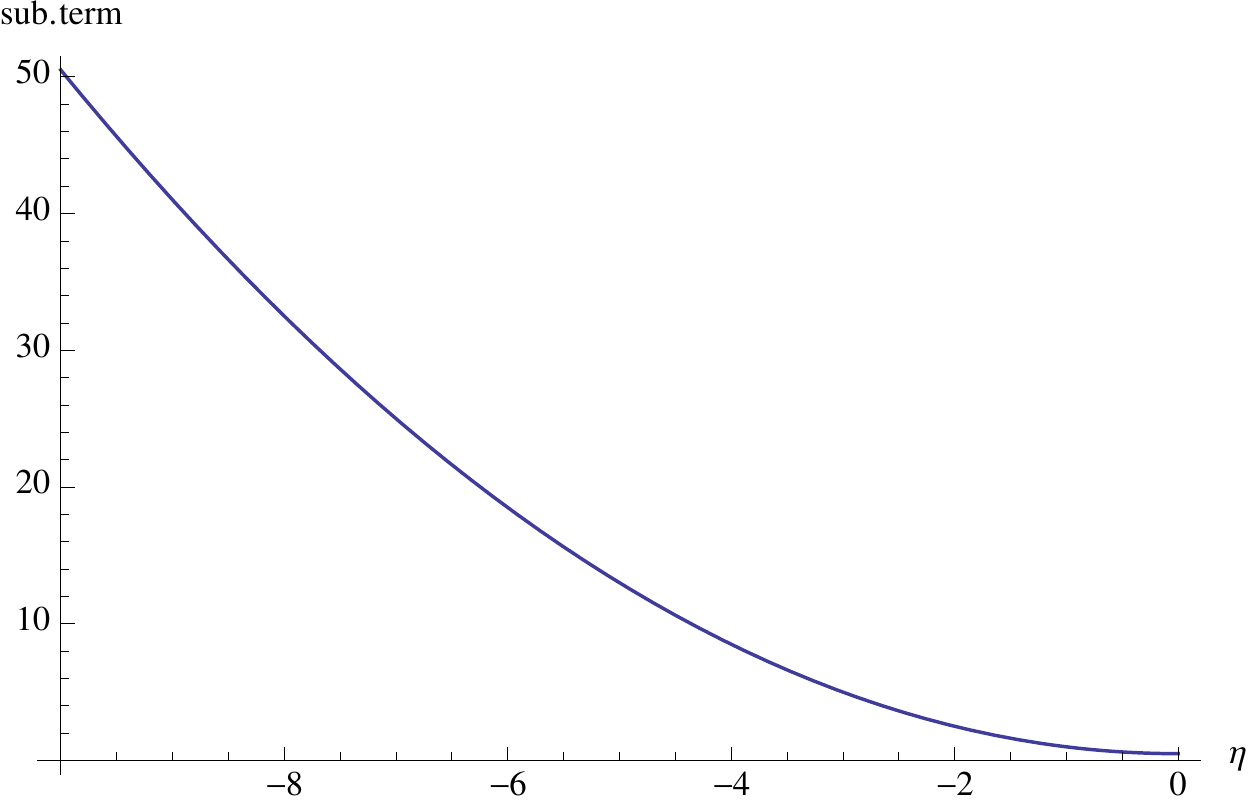}
    \caption{The time development of the subtraction term Eq. (\ref{eq:sub-dS}) in the case of de Sitter spacetime. We are assuming $\dot{\varphi}=1$ (constant) to plot. We also set $H=k=1$ because these constants affect only the overall magnitude of the power spectrum. The subtraction term does not become zero even at the late time of inflation era.}
    \label{fig:sub-dS}
\end{center} \end{figure}

Under the slow-roll condition, we assume that the second derivative of inflaton $\ddot{\varphi}$ is small. We can neglect the time dependence of $\dot{\varphi}$ in the denominator of Eq. (\ref{eq:sub-dS}) during the inflation, and find that the first term exponentially decays but the second term is constant in the time development.
It is shown in the Fig. (\ref{fig:sub-dS}).

Take the difference between the bare power spectrum after the horizon exit (in large scale limit) and the subtraction term,
\begin{equation}
\lim_{-k\eta \to 0}|\mathcal{R}_k (\eta)|_{\text{phys}}^2
=\frac{H^4}{2\dot{\varphi}^2 k^3}\biggr|_{\text{exit}}-\frac{H^4}{2\dot{\varphi}^2 k^3}~.
\label{eq:phys-dS}
\end{equation}
Because $H$ and $\dot{\varphi}$ is nearly constant during slow-roll inflation, the physical power spectrum in large scale limit is exceedingly suppressed by the subtraction term.
This result agrees with the prior research \cite{Parker:2007-1,Agullo:2010-1,Agullo:2011-1} which argue the amplitude of the fluctuation of not the comoving curvature but the inflaton itself: $\langle|\delta\varphi|^2\rangle_{\text{phys}}$ is also suppressed by the subtraction term during inflation.

\subsection{DBI inflation model} 
Next let us to see how the adiabatic subtraction term behaves in DBI inflation model \cite{Silverstein:2004-1}.
DBI inflation is motivated by string theory 
and it achieves the inflation by the ``D-cceleration'' mechanism.
The cosmological property of this model has been analyzed, see e.g. \cite{Silverstein:2004-1,Alishahiha:2004-1}.

The Lagrangian of DBI inflation model is given by
\begin{equation}
P(\varphi,X)=-\frac{1}{f_{\text{D}}(\varphi)}\left(\sqrt{1-2f_{\text{D}}(\varphi)X}-1\right)-V(\varphi)
\end{equation}
where $f_{\text{D}}(\varphi)$ is the (squared) warp factor and we redifine $X$ in terms of the proper time metric\footnote{We set the sring coupling $g_s =1$ because it only affects the overall magnitude of the power spectrum at large scale limit.}, i.e., $X=\frac{1}{2}\dot{\varphi}^2$.

The adiabatic subtraction term in this model can be calculated by using the general result Eq. (\ref{eq:mini-sub}) or Eq. (\ref{eq:sub-in-sr}).
\begin{equation}
|\mathcal{R}_k (\eta)|_{\text{sub}}^2
=\frac{1}{2z^2 c_s k}\left\{1+\left(\frac{aH}{c_s k}\right)^2 \left(1+\delta\epsilon+\delta c_s \right)\right\}
\tag{\ref{eq:sub-in-sr}}
\end{equation}

Let us calculate the speed of sound and $z$.
\begin{equation}
P,_X = \frac{1}{\sqrt{1-2f_{\text{D}}(\varphi)X}},
~P,_{XX}=\frac{f_{\text{D}}(\varphi)}{(1-2f_{\text{D}}(\varphi)X)^{\frac{3}{2}}}
\end{equation}
\begin{equation}
E+P=2XP,_X=\frac{2X}{\sqrt{1-2f_{\text{D}}(\varphi)X}}
\end{equation}
\begin{equation}
\therefore c_s^2\equiv\frac{P,_X}{2XP,_{XX}+P,_X}=1-2f_{\text{D}}(\varphi)X
\end{equation}
\begin{equation}
\therefore z^2 =\frac{a^4 (E+P)}{c_s^2 \mathcal{H}^2}
=\frac{2a^2}{H^2}\frac{X}{(1-2f_{\text{D}}(\varphi)X)^{\frac{3}{2}}}
\end{equation}
Note that both $c_s$ and $z$ do not depend on the potential $V(\varphi)$ in general (not limited to DBI model) because $E+P$ is derivative of $X$.
However, the behavior of $\varphi$ is determined by the $f_{D}(\varphi)$ and $V(\varphi)$.

\begin{figure}[h] \begin{center}
    \includegraphics[width=10cm]{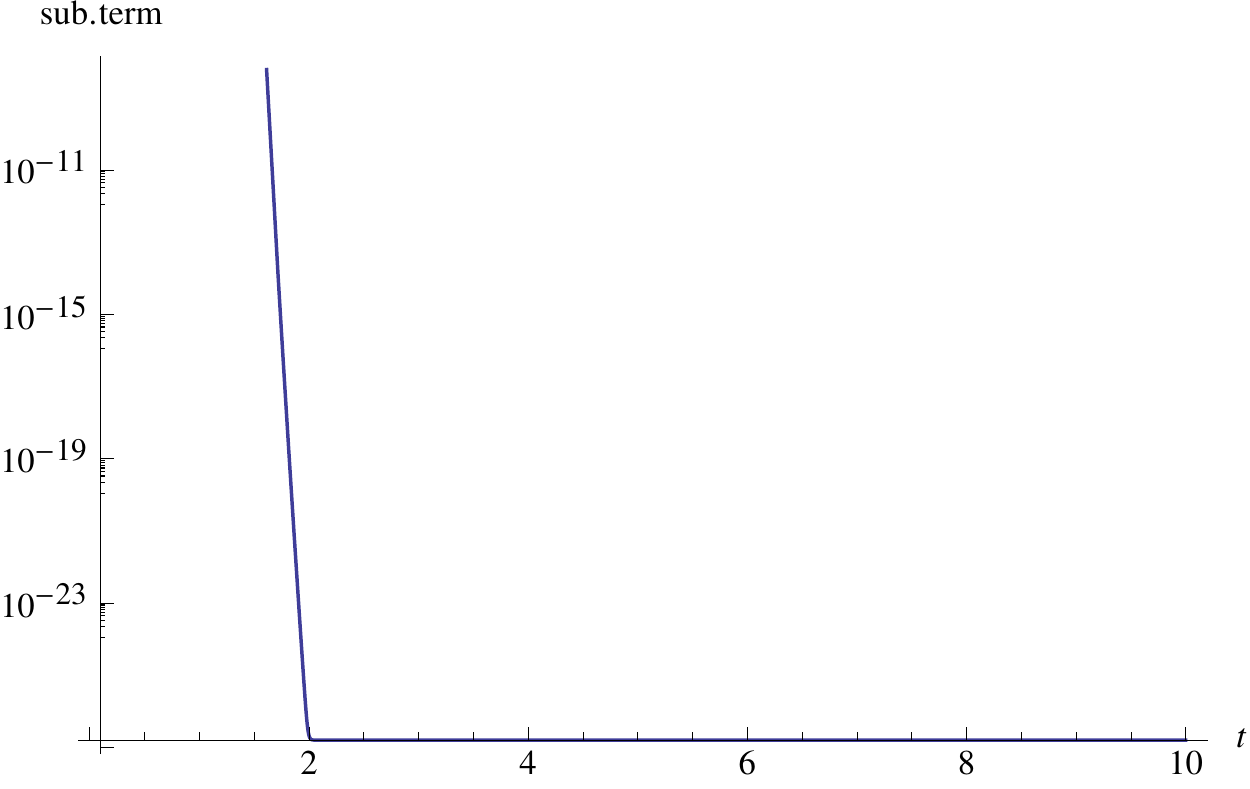}
    \caption{The time development of the subtraction term in the case of a DBI inflation model $f_{D}(\varphi)=\frac{\lambda}{\varphi^4}$, $V(\varphi)=m^2\varphi^2$.
    We set $k=1$, $\epsilon_D =0.01$ and $m=10^{-15}$ to plot.
    The subtraction term decreases very rapidly but the second order term which cancels the bare spectrum remains.
    }
    \label{fig:sub-DBI}
\end{center} \end{figure}

We can obtain the subtraction term in terms of $f_D (\varphi)$ and $X$ by substituting $z$ and $c_s$ into Eq. (\ref{eq:sub-in-sr}).
However, there are many choices of the two functions.
In this section, we assume a simple model which has $f_{D}(\varphi)=\lambda/\varphi^4$ and $V(\varphi)=m^2\varphi^2$.
$\lambda$ is the 't Hooft coupling and $m$ is the mass of the inflaton.
This model is analyzed in detail in \cite{Alishahiha:2004-1} and can achieve a power law inflation.

In this model, each function behaves at late time (but it is still inflation era) as following \cite{Alishahiha:2004-1}.
\begin{equation}
a(t)\rightarrow a_{\text{ini}} t^{1/\epsilon_{D}},\hspace{1em}
H\rightarrow\frac{1}{\epsilon_D t},\hspace{1em}
\varphi\rightarrow\frac{\sqrt{\lambda}}{t},\hspace{1em}
c_s\rightarrow\sqrt{\frac{3\lambda}{4}}\frac{1}{mt^2},\hspace{1em}
z\rightarrow z_{\text{ini}} t^{(2\epsilon_D +1)/\epsilon_D}
\label{eq:late-DBI}
\end{equation}
where $\epsilon_D$ is a slow-roll parameter.
\begin{equation}
\epsilon_D \equiv 2c_s \left(\frac{H,_{\varphi}}{H}\right)^2
\end{equation}
i.e. $c_s \ll 1$ is necessary to achieve accelerated expansion. 

$\epsilon_D$ can be calculated by using the Friedmann equation and Eq. (\ref{eq:late-DBI})
\begin{equation}
\epsilon_D
=\frac{3}{\sqrt{3m^2 \lambda +2\sqrt{3m^2 \lambda}}}
\approx \sqrt{\frac{3}{\lambda}}m^{-1}
\end{equation}
and it is approximately constant at late time.
Let us see the relationship between $\epsilon_D$ and the flow functions.
\begin{equation}
\epsilon_1 =-\frac{\dot{H}}{H^2}=\epsilon_D
\end{equation}
\begin{equation}
\delta_1 =-\frac{\dot{c_s}}{Hc_s}=-2\epsilon_D
\end{equation}
Therefore $\epsilon_D =\epsilon_1 = -\frac{1}{2}\delta_1$ and $\delta\epsilon$ and $\delta c_s$ are constants during the inflation.

Then we can also use the factors $\frac{1}{2z^2 c_s k}$ and $\frac{1}{2z^2 c_s k}\left(\frac{aH}{c_s k}\right)^2$ to see the time dependence of the subtraction term.
Substitute Eq. (\ref{eq:late-DBI}) into them and we obtain the following expression.
\begin{equation}
\frac{1}{2z^2 c_s k}=\frac{c_s}{4\epsilon_1 a^2 k}
=\frac{3}{8\epsilon_D^2 m^2 k} t^{-2(1+1/\epsilon_D)}
\end{equation}
\begin{equation}
\frac{1}{2z^2 c_s k}\left(\frac{aH}{c_s k}\right)^2
=\frac{H^2}{4\epsilon_D c_s k^3}
=\frac{m^2}{6 \epsilon_D^2 k^3}
=\text{const.}
\end{equation}
The zeroth adiabatic order factor becomes small, while the second term is constant.
This is also the same as the bare power spectrum of this model is given by (in the Bunch--Davies vacuum and in large scale limit)
\begin{equation}
|\mathcal{R}_k (\eta)|_{\text{bare}}^2 =\frac{1}{2\lambda\epsilon_D^4 k^3}
=\frac{m^2}{6 \epsilon_D^2 k^3}~.
\end{equation}
The time development is shown in the Fig. (\ref{fig:sub-DBI}).

Therefore the regularized power spectrum at the large scale is also strongly suppressed in this model.
\newpage
\section{The time development of the adiabatic subtraction terms after inflation}\label{chap:sub-dev-after}

As mentioned, we assume that the subtraction terms do not freeze out at horizon crossing.
The subtraction terms shrink after inflation in the canonical kinetic term case \cite{Urakawa:2009-1}.
In the non-canonical case, they also decrease if we assume the nearly scale invariant power spectrum \cite{Alinea:2015-1}.

Then, the following question comes to mind;
How small the subtraction terms become?
No one seems to check this quantitatively.
In this appendix, we see it via a very rough estimate for the following research.

We restrict ourselves to considering the models which have neither the non-canonical kinetic terms nor the coupling with the scalar curvature.
Also we assume that the scalar perturbation obeys the only one equation throughout the history of the universe, and we neglect the dark energy dominant era.

The second adiabatic order subtraction term is
\begin{equation}
\frac{1}{2z^2 k}\left(\frac{aH}{k}\right)^2 (1+\delta\epsilon)
\approx \frac{H^2}{4\epsilon_1 k^3}~.
\label{eq:2ndsub}
\end{equation}

For the simplicity of the formula, we set the magnitude of the dimensionless bare power spectrum to $1/8\pi^2$.
Then, to compare the magnitude of the subtraction terms with the bare one, we only have to see the time development of Eq. (\ref{eq:2ndsub}) for a specific mode.
Eq. (\ref{eq:2ndsub}) can be rewritten in terms of slow-roll parameters \cite{Urakawa:2009-1}.
\begin{equation}
\frac{H^2}{4\epsilon_1 k^3}=
\frac{1}{4k^3}\exp{\left(-\int^N_{N_*} d\tilde{N}(2\epsilon_1 (\tilde{N})+\epsilon_2 (\tilde{N}))\right)}
\label{eq:rewritten}
\end{equation}
where $N_*$ is the e-folding at the horizon exit.
Here we define the e-folding $N(a)\equiv\ln{(a/a_*)}$ with $a_0 =1$.

We need separate the range of integral to calculate this.
We consider three parts: inflation era (from $N_*$ to $N_{\text{end}}$), radiation dominant era (from $N_{\text{end}}$ to $N_{\text{eq}}$), and after that.

We assume the inflation ends abruptly at $N_{\text{end}}$.
The behavior of the subtraction terms during inflation era can be calculated (see Appendix \ref{chap:sub-dev}).
It has model dependence through the slow-roll parameters and e-folding, but is typically reduced to $10^{-1}$--$10^{-3}$.

After the inflation, the slow-roll parameters become large and the time behavior of the subtraction terms lose the inflational-model dependence (except for the sound speed).

In radiation dominant era, the integrand of Eq. (\ref{eq:rewritten}) is equal to $4$.
The subtraction terms become small $\exp{\left(-4(N_{\text{eq}}-N_{\text{end}})\right)}$.
$N_{\text{eq}}-N_{\text{end}}$ is depend on the temperature at the end of inflation era.
However, using the scale factor at the recombination and at the time the energy density of radiation and matter became equal, the subtraction terms are reduced at least to $10^{-2}$ from the value at the end of inflation.
Because
\begin{equation}
N_{\text{eq}}-N_{\text{end}} > \ln{\left(\frac{a_{\text{eq}}}{a_\text{CMB}}\right)}\approx 0.8
\end{equation}
where $a_\text{CMB}$ is the scale factor at the recombination.

The ambiguity of the calculation becomes trivial in matter dominant era.
In this era, the integrand of Eq. (\ref{eq:rewritten}) is equal to $3$ and
\begin{equation}
N_0 -N_{\text{eq}}=\ln{\left(\frac{1}{a_\text{eq}}\right)}\approx 6~.
\end{equation}
the subtraction terms are reduced to about $10^{-8}$.

Combining these results, the adiabatic subtraction terms are reduced at least to $10^{-11}$ by the present.
However, the power spectrum including subtraction terms should be estimated at the time it become ``classical''.

It is unknown when and how the power spectrum became classical.
It perhaps is in inflation era or radiation dominant era.
If the subtraction terms was not sufficiently small at the time, we may have the opportunity to observe the difference.

This calculation is very roughly.
We need more precise consideration including the effect by reheating and higher order quantum corrections.
\newpage\section{The subtraction terms in the Jordan frame}\label{chap:jordan-frame}
In this section, we calculate the explicit form of the subtraction terms in the Jordan frame by using the conformal transformation to see how the non-minimal coupling term affects on it.

The subtraction term has been derived in the section \ref{sec:conformal}.
\begin{equation}
|\widehat{\mathcal{R}} (\eta)|_{\text{sub}}^2=
\frac{1}{2\widehat{z}^2 \widehat{c}_s k}\left\{1+\frac{\widehat{z}''}{2\widehat{z}}\frac{1}{\widehat{c}_s^2 k^2}+\frac{1}{\widehat{c}_s^2 k^2}\left(\frac{1}{4}\frac{\widehat{c}''_s}{\widehat{c}_s}-\frac{3}{8}\frac{\widehat{c}'_s {}^2}{\widehat{c}_s^2}\right)\right\}
\tag{\ref{sub:einstein}}
\end{equation}
where
\begin{equation}
\widehat{z}^2=2\left\{3e^{2\theta}\left(\frac{\mathcal{H}}{\theta'}-1\right)^2+\frac{a^4 \Sigma}{\theta'^2}\right\}
\tag{\ref{eq:zcon}}
\end{equation}
\begin{equation}
\widehat{c}_s^2=\frac{e^{2\theta}\left(1-\frac{\theta''}{\theta' {}^{2}}\right)}{3e^{2\theta}\left(\frac{\mathcal{H}}{\theta'}-1\right)^2+\frac{a^4 \Sigma}{\theta'^2}}=\frac{\theta'^2-\theta''}{3(\mathcal{H}-\theta')^2+\frac{a^4 \Sigma}{e^{2\theta}}}
=\frac{2\widehat{\epsilon}_1 \widehat{a}^2}{\widehat{z}^2}~.
\tag{\ref{eq:ccon}}
\end{equation}

Let us rewrite $\widehat{z}$ in terms of $z$ which is the variables in the minimal case.
\begin{equation}
\widehat{z}
=\frac{\mathcal{H}}{\theta'}\sqrt{6e^{2\theta}\left(1-\frac{\theta'}{\mathcal{H}}\right)^2 +2\frac{a^4 \Sigma}{\mathcal{H}^2}}
=\frac{\mathcal{H}}{\theta'}\sqrt{6e^{2\theta}\left(1-\frac{\theta'}{\mathcal{H}}\right)^2 +z^2}
\end{equation}

To derive the explicit form of the subtraction terms in the Jordan frame, we need the second derivative of $\widehat{z}$ and $\widehat{c}_s$.
We can use some convenient calculations as below.
\begin{equation}
\widehat{z}'=\frac{1}{2\widehat{z}}\frac{d}{d\eta}\widehat{z}^2,\hspace{1em}
\frac{\frac{d}{d\eta}\widehat{z}'^2}{\frac{d}{d\eta}\widehat{z}^2}
=\frac{2\widehat{z}'\widehat{z}''}{2\widehat{z}\widehat{z}'}
=\frac{\widehat{z}''}{\widehat{z}}
\label{eq:derivative}
\end{equation}
\begin{equation}
\frac{d}{d\eta}\left(\frac{AB\dots}{CD\dots}\right)^2
=2\left(\frac{AB\dots}{CD\dots}\right)^2 \left(\frac{A'}{A}+\frac{B'}{B}+\dots-\frac{C'}{C}-\frac{D'}{D}-\dots\right)
\end{equation}

\subsection*{Derivatives of $\widehat{z}$}
Let us calculate the derivatives of $\widehat{z}$ first.
For simplicity, define a function $\alpha$,
\begin{equation}
\alpha^2 \equiv 6e^{2\theta}\left(1-\frac{\theta'}{\mathcal{H}}\right)^2,
\end{equation}
This function approaches to zero in minimal coupling limit bacause $\theta'$ is equal to $\mathcal{H}$ in the case of minimal coupling model.

Then $\widehat{z}^2$ becomes
\begin{equation}
\widehat{z}^2 = \left(\frac{\mathcal{H}}{\theta'}\right)^2 (z^2 +\alpha^2)
\label{zhatal}
\end{equation}
and
\begin{eqnarray}
\frac{d}{d\eta}\widehat{z}^2 &=&
2\left(\frac{\mathcal{H}}{\theta'}\right)^2 z^2 \left(\frac{\mathcal{H}'}{\mathcal{H}}+\frac{z'}{z}-\frac{\theta''}{\theta'}\right) +2\left(\frac{\mathcal{H}}{\theta'}\right)^2 \alpha^2 \left(\frac{\mathcal{H}'}{\mathcal{H}}+\frac{\alpha'}{\alpha}-\frac{\theta''}{\theta'}\right)\nonumber\\
&=& 2\left(\frac{\mathcal{H}}{\theta'}\right)^2\left\{(z^2 +\alpha^2)\left(\frac{\mathcal{H}'}{\mathcal{H}}-\frac{\theta''}{\theta'}\right)+zz'+\alpha\alpha'\right\}~.
\end{eqnarray}
Using the relation Eq. (\ref{eq:derivative}),
\begin{eqnarray}
\widehat{z}'^2 &=& \frac{1}{4\widehat{z}^2}\left(\frac{d}{d\eta}\widehat{z}^2\right)^2\nonumber\\
&=& \frac{4\left(\frac{\mathcal{H}}{\theta'}\right)^4\left\{(z^2 +\alpha^2)\left(\frac{\mathcal{H}'}{\mathcal{H}}-\frac{\theta''}{\theta'}\right)+zz'+\alpha\alpha'\right\}^2}{4\left(\frac{\mathcal{H}}{\theta'}\right)^2 (z^2 +\alpha^2)}\nonumber\\
&=&\frac{1}{z^2 +\alpha^2}\left(\frac{\mathcal{H}}{\theta'}\right)^2\left\{zz'+\alpha\alpha'+(z^2 +\alpha^2)\left(\frac{\mathcal{H}'}{\mathcal{H}}-\frac{\theta''}{\theta'}\right)\right\}^2~.
\end{eqnarray}

Again, to simplify this, define $\beta$ vanishing in minimal coupling limit
\begin{equation}
\beta\equiv\alpha\alpha'+(z^2 +\alpha^2)\left(\frac{\mathcal{H}'}{\mathcal{H}}-\frac{\theta''}{\theta'}\right)
\end{equation}
so that
\begin{equation}
\widehat{z}'^2=\frac{1}{z^2 +\alpha^2}\left(\frac{\mathcal{H}}{\theta'}\right)^2\left(zz'+\beta\right)^2~.
\end{equation}
Similarly,
\begin{equation}
\frac{d}{d\eta}\left(\widehat{z}'^2\right)
=\frac{2}{z^2 +\alpha^2}\left(\frac{\mathcal{H}}{\theta'}\right)^2\left(zz'+\beta\right)^2 \left\{\left(\frac{\mathcal{H}'}{\mathcal{H}}-\frac{\theta''}{\theta'}\right)-\frac{zz'+\alpha\alpha'}{z^2+\alpha^2}+\frac{z'^2 +zz''+\beta'}{zz'+\beta}\right\}~,
\end{equation}
\begin{eqnarray}
\frac{\widehat{z}''}{\widehat{z}}=\frac{\frac{d}{d\eta}\widehat{z}'^2}{\frac{d}{d\eta}\widehat{z}^2}
&=&\frac{\frac{2}{z^2 +\alpha^2}\left(\frac{\mathcal{H}}{\theta'}\right)^2\left(zz'+\beta\right)^2 \left\{\left(\frac{\mathcal{H}'}{\mathcal{H}}-\frac{\theta''}{\theta'}\right)-\frac{zz'+\alpha\alpha'}{z^2+\alpha^2}+\frac{z'^2 +zz''+\beta'}{zz'+\beta}\right\}}{2\left(\frac{\mathcal{H}}{\theta'}\right)^2\left\{(z^2 +\alpha^2)\left(\frac{\mathcal{H}'}{\mathcal{H}}-\frac{\theta''}{\theta'}\right)+zz'+\alpha\alpha'\right\}}\nonumber\\
&=&\left(\frac{zz'+\beta}{z^2+\alpha^2}\right)^2 \frac{\left(\frac{\mathcal{H}'}{\mathcal{H}}-\frac{\theta''}{\theta'}\right)-\frac{zz'+\alpha\alpha'}{z^2+\alpha^2}+\frac{z'^2 +zz''+\beta'}{zz'+\beta}}{\left(\frac{\mathcal{H}'}{\mathcal{H}}-\frac{\theta''}{\theta'}\right)+\frac{zz'+\alpha\alpha'}{z^2+\alpha^2}}~.
\end{eqnarray}

This is the expression of coefficient function of the second term of Eq. (\ref{sub:einstein}) in the Jordan frame.

\subsection*{Derivatives of $\widehat{c}_s$}
Next, we need the expression of the derivatives of $\widehat{c}_s$ in the Jordan frame.
By definition,
\begin{equation}
\widehat{c}_s^2 =\frac{2\widehat{\epsilon}_1 \widehat{a}^2}{\widehat{z}^2}~.
\end{equation}
Using Eq.(\ref{zhatal}) and Eq. (\ref{eq:hatep}),
\begin{equation}
\frac{1}{\widehat{c}_s^2}=\frac{1}{2}\left(\frac{\mathcal{H}}{\theta'}\right)^2\frac{z^2 +\alpha^2}{1-\frac{\theta''}{\theta'}}e^{-2\theta}~.
\end{equation}

To see the difference from minimal coupling model, rewrite $\widehat{\epsilon}_1$.
\begin{equation}
\widehat{\epsilon}_1 =1-\frac{\theta''}{\theta'^2}
=1-\frac{\mathcal{H}'}{\mathcal{H}^2}+\left(\frac{\mathcal{H}'}{\mathcal{H}^2}-\frac{\theta''}{\theta'^2}\right)
\end{equation}
where we define $\gamma$ as
\begin{equation}
\gamma\equiv\frac{\mathcal{H}'}{\mathcal{H}^2}-\frac{\theta''}{\theta'^2}
\end{equation}
and it vanishes in the minimal coupling limit.
$1-\frac{\mathcal{H}'}{\mathcal{H}^2}=\epsilon_1$ is the first Hubble flow function of the minimal coupling model.

Then we get
\begin{equation}
\frac{1}{\widehat{c}_s^2}=\frac{1}{2}\left(\frac{\mathcal{H}}{\theta'}\right)^2\frac{z^2 +\alpha^2}{\epsilon_1 +\gamma}e^{-2\theta}
\end{equation}
and calculate the derivatives.
\begin{eqnarray}
&&\frac{d}{d\eta}\widehat{c}_s^2 =\frac{d}{d\eta}\left[2e^{2\theta}\left(\frac{\theta'}{\mathcal{H}}\right)^2 \frac{\epsilon_1 +\gamma}{z^2 +\alpha^2}\right]\nonumber\\
&=&2e^{2\theta}\left(\frac{\theta'}{\mathcal{H}}\right)^2 \frac{\epsilon_1 +\gamma}{z^2 +\alpha^2}\left\{2\theta'+2\left(\frac{\theta''}{\theta'}-\frac{\mathcal{H'}}{\mathcal{H}}\right)+\frac{\epsilon'_1 +\gamma'}{\epsilon_1 +\gamma}-\frac{2(zz'+\alpha\alpha')}{z^2 +\alpha^2}\right\}
\end{eqnarray}

Define $\kappa$ as
\begin{equation}
\kappa\equiv2\theta'+2\left(\frac{\theta''}{\theta'}-\frac{\mathcal{H'}}{\mathcal{H}}\right)+\frac{\epsilon'_1 +\gamma'}{\epsilon_1 +\gamma}-\frac{2(zz'+\alpha\alpha')}{z^2 +\alpha^2}
\label{eq:kappa}
\end{equation}
so that
\begin{equation}
\frac{d}{d\eta}\widehat{c}_s^2 = \widehat{c}_s^2 \kappa~.
\end{equation}
$\kappa$ approaches to $2\frac{c'_s}{c_s}$ in minimal coupling limit.

Therefore
\begin{equation}
\widehat{c}'_s =\frac{1}{2\widehat{c}_s}\frac{d}{d\eta}\widehat{c}_s^2
=\frac{1}{2}\widehat{c}_s \kappa
\end{equation}
\begin{equation}
\frac{d}{d\eta}(\widehat{c}'_s)^2 =\frac{1}{4}\kappa^2 \frac{d}{d\eta}\widehat{c}_s^2 +\frac{1}{4}\widehat{c}_s^2 \frac{d}{d\eta}\kappa^2
=\frac{1}{4}\widehat{c}_s^2 \kappa^3+\frac{1}{4}\widehat{c}_s^2 \frac{d}{d\eta}\kappa^2
\end{equation}
\begin{equation}
\frac{\widehat{c}''_s}{\widehat{c}_s}=\frac{\frac{d}{d\eta}(\widehat{c}'_s )^2}{\frac{d}{d\eta}\widehat{c}_s^2}
=\frac{1}{4}\kappa^2+\frac{1}{4}\frac{1}{\kappa}\frac{d}{d\eta}\kappa^2
=\frac{1}{4}\kappa^2 +\frac{1}{2}\kappa'
\end{equation}
\begin{equation}
\frac{\widehat{c}'_s {}^2}{\widehat{c}_s^2}=\frac{\frac{1}{4}\widehat{c}_s^2 \kappa^2}{\widehat{c}_s^2}=\frac{1}{4}\kappa^2
\end{equation}
\begin{equation}
\therefore \frac{1}{4}\frac{\widehat{c}''_s}{\widehat{c}_s}-\frac{3}{8}\frac{\widehat{c}'_s {}^2}{\widehat{c}_s^2}
=\frac{1}{8}\left(\kappa'-\frac{1}{4}\kappa\right)
\label{eq:kappas}
\end{equation}
where $\kappa$ is defined as Eq. (\ref{eq:kappa}).

\subsection*{Results}
Substitute the expressions to Eq.(\ref{sub:einstein}),
we get
\begin{eqnarray}
|\widehat{\mathcal{R}} (\eta)|_{\text{sub}}^2&=&
\frac{1}{2k}\frac{1}{\sqrt{2(\epsilon_1 +\gamma)(z^2 +\alpha^2)}}\left(\frac{\theta'}{\mathcal{H}}\right)e^{-\theta}\left[1+
\frac{1}{4k^2}\left(\frac{\mathcal{H}}{\theta'}\right)^2\frac{z^2+\alpha^2}{\epsilon_1 +\gamma}e^{-2\theta}\right.\\
&\times&\left.\left\{
\left(\frac{zz'+\beta}{z^2+\alpha^2}\right)^2 \frac{\left(\frac{\mathcal{H}'}{\mathcal{H}}-\frac{\theta''}{\theta'}\right)-\frac{zz'+\alpha\alpha'}{z^2+\alpha^2}+\frac{z'^2 +zz''+\beta'}{zz'+\beta}}{\left(\frac{\mathcal{H}'}{\mathcal{H}}-\frac{\theta''}{\theta'}\right)+\frac{zz'+\alpha\alpha'}{z^2+\alpha^2}}
+\frac{1}{4}\left(\kappa'-\frac{1}{4}\kappa\right)
\right\}\right]~.\nonumber
\end{eqnarray}

We have to calculate the derivatives of $\alpha$, $\beta$, and $\kappa$ to obtain the final expression.
However, the expression is very complicated and there seems to be no further use in continuing in this fashion. It is actually enough to see that the coincidence of the expression in the minimal coupling limit, and the results definitely coincide. Although our expression of the adiabatic subtraction terms for ``general'' (non-minimal) k-inflation depends on the non-minimal coupling term in a complex manner, it includes the expression for the usual (minimal) k-inflation as a special case.


\label{Bibliography}
\bibliographystyle{unsrt}  

\bibliography{Bibliography}
\end{document}